\documentclass[9pt,twocolumn,twoside]{pnas-new} 

\templatetype{pnasresearcharticle}
\setboolean{displaywatermark}{false}

\usepackage{amsmath,amssymb,amsfonts}
\usepackage{graphicx}
\usepackage{textcomp}
\usepackage{longtable}

\title{Moving beyond simulation: data-driven quantitative photoacoustic imaging using tissue-mimicking phantoms}

\leadauthor{Gr\"ohl}

\author[ a,b]{Janek Gr\"ohl}
\author[ a,b]{Thomas R. Else}
\author[ a,b,c]{Lina Hacker}
\author[ a,b]{Ellie V. Bunce}
\author[ a,b]{Paul W. Sweeney}
\author[ a,b]{Sarah E. Bohndiek}
\affil[a]{Cancer Research UK Cambridge Institute, University of Cambridge, Cambridge, U.K.}
\affil[b]{Department of Physics, University of Cambridge, Cambridge, U.K.}
\affil[c]{L.H. is now at the Department of Oncology, University of Oxford, Oxford, UK.}

\correspondingauthor{Correspondence: \{jmg236, seb53\}@cam.ac.uk}


\begin{abstract}
Accurate measurement of optical absorption coefficients from photoacoustic imaging (PAI) data would enable direct mapping of molecular concentrations, providing vital clinical insight. The ill-posed nature of the problem of absorption coefficient recovery has prohibited PAI from achieving this goal in living systems due to the domain gap between simulation and experiment. To bridge this gap, we introduce a collection of experimentally well-characterised imaging phantoms and their digital twins. This first-of-a-kind phantom data set enables supervised training of a U-Net on experimental data for pixel-wise estimation of absorption coefficients. We show that training on simulated data results in artefacts and biases in the estimates, reinforcing the existence of a domain gap between simulation and experiment. Training on experimentally acquired data, however, yielded more accurate and robust estimates of optical absorption coefficients. We compare the results to fluence correction with a Monte Carlo model from reference optical properties of the materials, which yields a quantification error of approximately 20\%. Application of the trained U-Nets to a blood flow phantom demonstrated spectral biases when training on simulated data, while application to a mouse model highlighted the ability of both learning-based approaches to recover the depth-dependent loss of signal intensity. We demonstrate that training on experimental phantoms can restore the correlation of signal amplitudes measured in depth. While the absolute quantification error remains high and further improvements are needed, our results highlight the promise of deep learning to advance quantitative PAI.
\end{abstract}

\begin{document} 
\maketitle
\thispagestyle{fancy}
\ifthenelse{\boolean{shortarticle}}{\ifthenelse{\boolean{singlecolumn}}{\abscontentformatted}{\abscontent}}{}

\section{Introduction}
\label{sec:introduction}

Photoacoustic (PA) imaging (PAI) has the unique ability to provide high-resolution optical-contrast molecular imaging up to several centimetres deep in tissue with  potential clinical application areas ranging from diagnosis of breast cancer~\cite{manohar2019current} or cardiovascular disease~\cite{dean2013volumetric}, to grading of skin~\cite{chuah2017structural,aguirre2017precision} or inflammatory diseases~\cite{knieling2017multispectral}. PAI transforms absorbed laser energy into acoustic sound waves~\cite{xu2006photoacoustic,beard2011biomedical} that can be measured as time-series pressure data $p(t)$ using acoustic detectors ~\cite{jaffe1965piezoelectric,vaughan2017fabry}.

A key aim of PAI is estimation of absolute molecular concentrations from images acquired at multiple wavelengths ~\cite{cox2009challenges,cox2012quantitative}. To achieve this goal, two inverse problems have to be solved: the \textit{acoustic inverse problem}, which refers to the reconstruction of the initial pressure distribution $p_0$ from $p(t)$ and the \textit{optical inverse problem}, which refers to the estimation of the optical absorption coefficient $\mu_a$ from $p_0$. Reconstructing $p_0$ from $p(t)$ requires an inverse acoustic operator, denoted as $A^{-1}$ (an image reconstruction algorithm). To obtain accurate reconstructions, the acoustic properties of the tissue, $\theta$, and the characteristics of the acoustic detector, $\epsilon$, must be taken into consideration: $ p_0 \approx A^{-1}(p(t, \epsilon), \theta, \epsilon).$

Once reconstructed, determining $\mu_a$ from $p_0$, for a given wavelength $\lambda$ at position $\textbf{x}$, must deal with the fact that $p_0$ depends not only on the local absorption coefficient $\mu_a$, but also the temperature ($T$)-dependent Gr\"uneisen parameter $\Gamma$ and the light fluence $\phi$, which is a function of both $\mu_a$ and the optical scattering coefficient $\mu_s$: $ p_0(\textbf{x}, \lambda, T) = \Gamma(\textbf{x},T)\cdot\mu_a(\textbf{x},\lambda)\cdot\phi(\textbf{x},\lambda, \mu_a(\lambda), \mu_s(\lambda)).$

The spatial distribution of $\phi$ and $\Gamma$ in living subjects are typically unknown but must be accounted for to accurately determine $\mu_a$. $\Gamma$ is often assumed constant~\cite{cox2012quantitative}. $\phi$ could be accounted for by iterative methods, using fixed-point iteration schemes assuming \textit{a priori} known $\mu_s$ ~\cite{cox2006two} or variational approaches, allowing a range of constraints and priors~\cite{tarvainen2013bayesian}. To date, these methods have been restricted to simulated data, except when including a calibrated imaging target into the medium~\cite{buchmann2020quantitative}. Deep learning approaches have been proposed to solve both the acoustic ~\cite{hauptmann2018model,antholzer2019deep} and optical inverse problems~\cite{cai2018end,grohl2018confidence,chen2020deep,li2022deep} in PAI. Iterative or variational methods typically work well with simulated data, but do not achieve convincing results on experimental data~\cite{grohl2021deep}. A key limitation for application in living subjects is that the optical and acoustic properties are unknown, making validation of quantitative PAI methods extremely difficult. One indirect possibility to validate these methods is to analyse the performance of methods that use the estimated $\mu_a$ as an input, for example, semantic image segmentation~\cite{schellenberg2022photoacoustic,dreher2023unsupervised}, or oximetry~\cite{grohl2021learned}.

\begin{figure}[h!tb]
    \centering
    \includegraphics[width=\columnwidth]{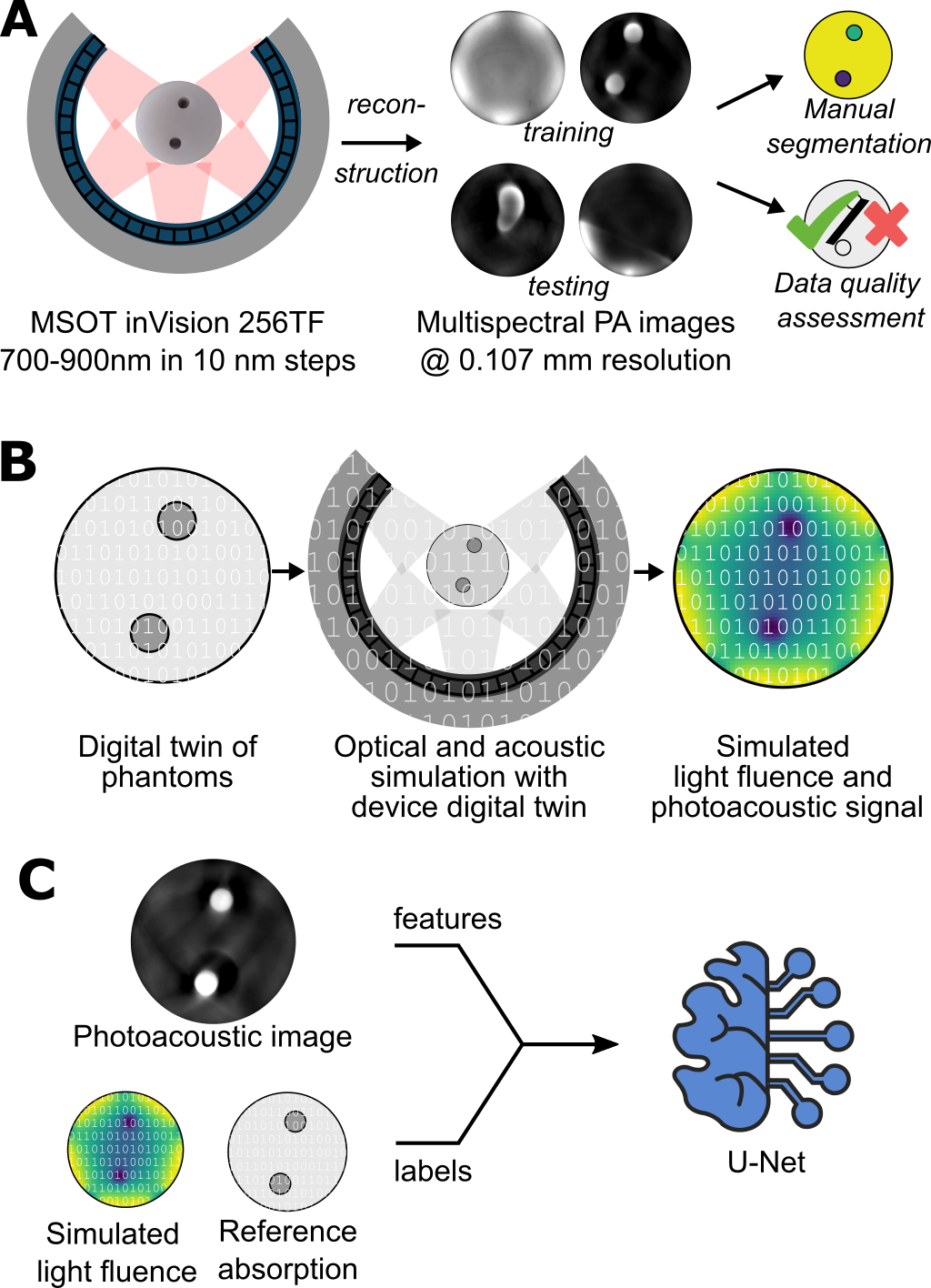}
    \caption{\textbf{Supervised deep learning approach.} A library of tissue-mimicking phantoms with known optical properties were subjected to: (\textbf{A}) Experimental photoacoustic imaging (PAI), followed by delay-and-sum reconstruction, manual image quality assessment, and segmentation; and (\textbf{B}) Computation of light fluence and time series pressure using digital twins of the phantom and imaging system. (\textbf{C}) Paired experimental and computational data were used to train a supervised deep learning model to estimate optical absorption coefficients.}
    \label{fig:methods1}
\end{figure}

Here, we bridge the gap from simulated to experimental training data by fabricating a collection of well-characterised co-polymer-in-oil tissue-mimicking phantoms~\cite{hacker2021copolymer} covering a biologically relevant range of $\mu_a$ and $\mu_s$, to enable the use of supervised deep learning for the optical inverse problem of PAI (Fig.~\ref{fig:methods1}). We create digital phantom twins by characterising optical properties with a double integrating sphere (DIS) system and train U-Nets~\cite{ronneberger2015u} to estimate $\mu_a$ on matching simulated and experimental PA images. We apply the trained U-Nets to a held-out set of test phantoms and compare the results to fluence correction with a Monte Carlo model based on the DIS reference measurements. Importantly, we then explore the potential of the methods on pre-clinical data from living subjects.

\section{Methods}
\label{sec:methods}

\subsection{Tissue mimicking phantoms}

A total of 137 cylindrical phantoms with a diameter of 27.5\,mm, a height of 65-80\,mm, and an approximate volume of 40-50\,mL were fabricated based on a previously published protocol\cite{hacker2021copolymer}. Briefly, 83.80\,g mineral oil (Sigma Aldrich-330779-1L) was used as a base. 25.14\,g polystyrene-block-poly(ethylene-ran-butylene)-block-polystyrene (SEBS, Sigma Aldrich 200557-250G) as the polymer and 2.09\,g butylated hydroxytoluene (HT, Sigma Aldrich W218405-1KG-K) as an antioxidant was added to the mineral oil base, which was heated to 150$^\circ$C and poured into a mould (for details see \cite{hacker2021copolymer}, \cite{hacker2022thesis}, and Supplementary Figure 1). Different concentrations of titanium dioxide (TiO$_2$, Sigma Aldrich 232033-100g) and alcohol-soluble nigrosin (Sigma Aldrich 211680-100G) were used to tune $\mu_s'$ and $\mu_a$ respectively. Each phantom consisted of a background cylinder into which imaging targets (referred to as inclusions) were added. Addition of the inclusions was performed using either careful pouring (diameter $>$ 6\,mm) or via a vacuum pump. Optical properties were sampled from the ranges $\mu_a$ = 0.05\,cm$^{-1}$ -- 4.0\,cm$^{-1}$ and $\mu_s'$ = 5\,cm$^{-1}$ -- 15\,cm$^{-1}$, defined at a reference wavelength of 800\,nm. The phantoms featured various inclusion patterns with distinct $\mu_a$ and $\mu_s'$ combinations in both the inclusions and the backgrounds (detailed in Supplementary Figure 2 and Supplementary Table 1).

\subsection{Optical property characterisation}

For each phantom material, rectangular samples with a length of 5.9 cm, a width of 1.8 cm, and a thickness ranging between 2\,mm and 4\,mm were prepared for optical characterisation. Sample thicknesses were determined at five distinct locations using digital vernier calipers. We used an in-house double integrating sphere (DIS) system~\cite{hacker2021copolymer} based on the system developed by Pickering et al.~\cite{pickering1993double} to determine $\mu_a$ and $\mu_s'$ of the phantom material in a wavelength range of 600 to 950 nm. The recorded transmittance and reflectance values were normalised and analysed with the inverse adding doubling (IAD) algorithm~\cite{prahl2011everything} to estimate the optical properties of the material. The anisotropy (g) and refractive index (n) were assumed to be g = 0.7 and n = 1.4~\cite{jones2018stability}.

\subsection{Photoacoustic imaging and data processing}

The phantoms were imaged using a pre-clinical PAI system (MSOT inVision-256TF, iThera Medical GmbH, Munich, Germany) described in detail elsewhere~\cite{joseph2017evaluation}. We imaged in the wavelength range from 700\,nm to 900\,nm in 10\,nm steps using 10 frame averaging. The phantoms were mounted in the imaging system using a custom 3D-printed polylactic acid phantom holder printed with the Prusa i3 MK3S+ (Prusa Research, Czechia) and immersed in deionized water for acoustic coupling. The phantoms were imaged at different cross-sections in 2\,mm steps along a 44\,mm-long section. The temperature of the water bath was set to 25\,$^{\circ}$C.

Images were reconstructed using the delay-and-sum algorithm with the PATATO toolbox~\cite{patato2023} with a bandpass filter between 5\,KHz and 7\,MHz, application of a Hilbert transform, as well as time interpolation by a factor of three and detection element interpolation by a factor of two. The images were reconstructed into a 300$\times$300 pixel grid with a field of view of 32$\times$32\,mm. To enable compatibility with the U-Net, images were cropped to 288$\times$288 pixels without changing the resolution.

\subsection{Data quality assurance}
A wide range of artefacts can corrupt the phantom preparation and imaging process when creating a wide range of inclusion geometries. Sources of artefact include: unsolved polymers; material inhomogeneities; air bubbles; yellowing from oxidation; and acoustic reflections from the 3D-printed phantom holder. After image reconstruction, manual quality control was performed for each tomographic cross-section to check for the presence of artefacts that impacted data quality. Phantoms were only included in subsequent analysis if at least one cross-sectional slice contained no artefacts. Using this process, 35 of the 137 phantoms were excluded.

\subsection{Synthetic imaging with digital twins}
We manually segmented the cross-sectional images using the Medical Imaging Interaction Toolkit (MITK)~\cite{wolf2004medical} into segmentation classes of background and target inclusions, which were used for creating the digital phantom twins.
 
We implemented a digital twin of the PAI device including both the illumination and detection geometry. The light source consists of five fibre bundle pairs radially distributed around the object and was implemented as a custom illumination in Monte Carlo eXtreme (MCX)~\cite{fang2009monte}. The implementation of this system can be found in {\color{blue}{https://github.com/jgroehl/mcx}} (last accessed 04/04/2023). The detection geometry consists of 256 toroidally focussed transducer elements covering a 270\,$^{\circ}$ field of view around the imaging target. The arrangement of the transducers was implemented in the k-Wave toolbox~\cite{treeby2010k}.

Each segmentation class was assigned the optical properties determined by the DIS measurements for the sample material at 21 wavelengths from 700\,nm to 900\,nm in 10\,nm steps. The background medium was assumed to be pure water~\cite{hale1973optical}, as this is the acoustic couplant used in the imaging system. We simulated the forward models using the SIMPA toolkit~\cite{grohl2022simpa}. The light distribution was simulated in 3D using the MCX Monte Carlo model and the propagation of sound waves was simulated in 2D using a custom k-Wave script. For computational efficiency, we limited the Monte Carlo model to $10^7$ photons for each simulation run and only simulated the acoustic model in 2D assuming an infinitely symmetric initial pressure outside of the imaging plane. Despite all assumptions, the simulations correlate well with the experiment (R=0.94, Supplementary Figure 3).

\subsection{Training and held-out test data.} 

The 102 phantoms that passed the quality assurance step were separated into a training set with 84 phantoms and a test set with 18 phantoms, roughly corresponding to the typically used 80/20 data split in deep learning. Since we imaged at 21 distinct wavelengths, we had 1764 training and 378 test data images. Test data was derived from a held out set of phantoms which did not contain base material $\mu_a$ and $\mu_s'$ value pairs that were used in the training set. Furthermore, ten of the test phantoms contained inclusion geometries not present in the training data.

\subsection{Absorption estimation methods}

\begin{figure}[h!tb]
    \centering
    \includegraphics[width=\columnwidth]{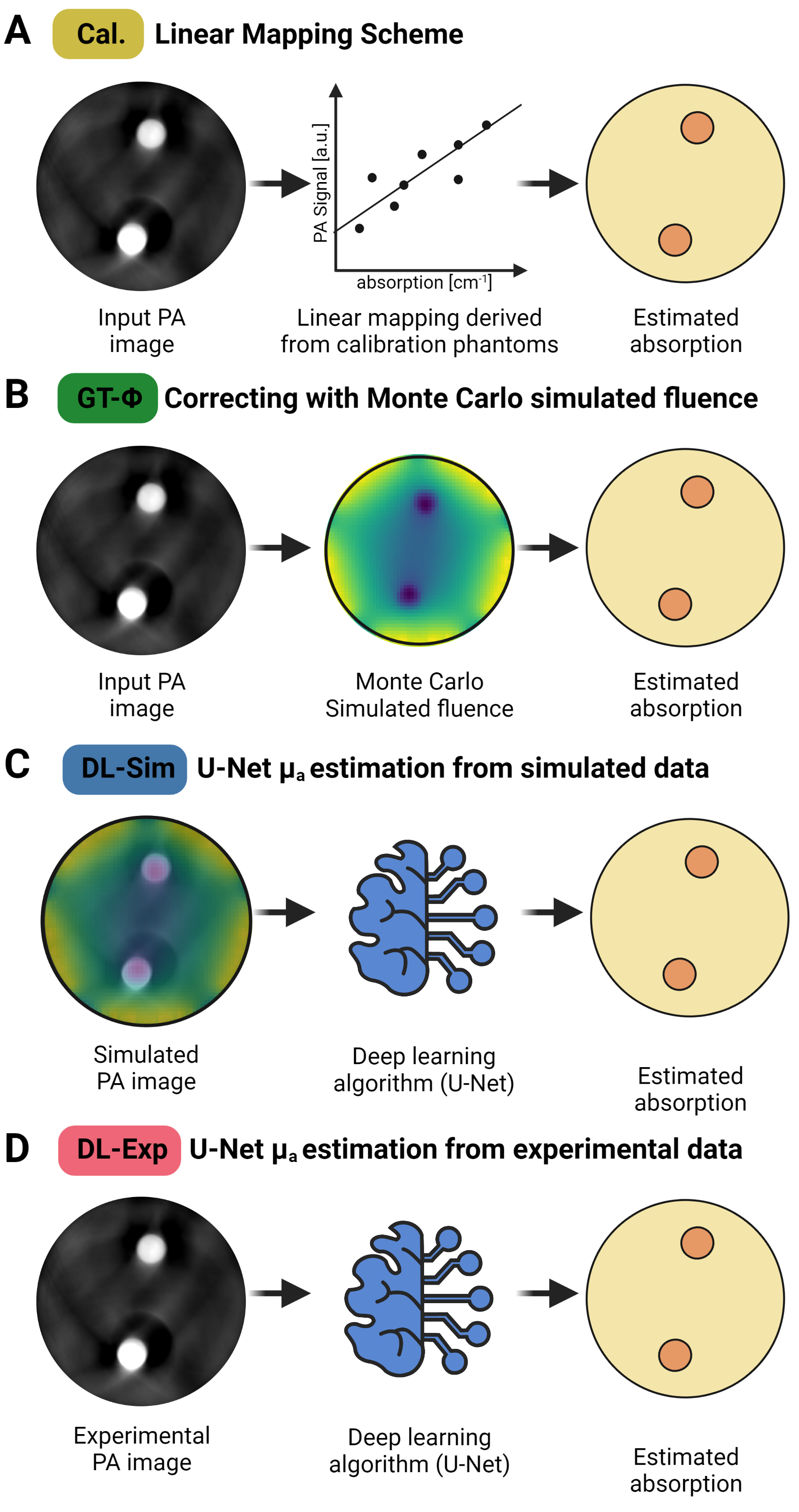}
    \caption{\textbf{Optical absorption estimation methods.} (\textbf{A}) The baseline model (Cal.) is a linear mapping from the arbitrary units PA image to reference absorption units. (\textbf{B}) Adding complexity, the simulated light fluence is used to correct PA images (GT-$\Phi$) before a linear mapping is applied. (\textbf{C}) For supervised deep learning, a U-Net is trained on simulated input data to estimate the optical absorption based on the input PAI data (DL-Sim) and (\textbf{D}) trained on experimentally acquired data (DL-Exp).}
    \label{fig:methods2}
\end{figure}

We applied four methods for calibration of the optical absorption coefficient $\mu_a$ against the PAI data (Fig.~\ref{fig:methods2}).

\textbf{Linear calibration} was chosen as a baseline model. To account for depth-dependent attenuation, we selected the brightest 2\% of pixels in the background material of the training phantoms (i.e. the rim) and applied a linear regression to relate the arbitrary units PA signals in these pixels to the reference optical absorption (Cal., Fig.~\ref{fig:methods2}A), giving $\mu_a = \frac{1}{1485}(\hat{p_0} - 313)$, R=0.94.

\textbf{Fluence correction with reference Monte Carlo simulations} was performed as a gold standard method. Monte Carlo simulations of light fluence were made based on the reference absorption and scattering values obtained by the DIS measurements (GT-$\phi$, Fig.~\ref{fig:methods2}B). The resulting linear fit on the fluence corrected data gave $\mu_a = \frac{1}{8801}(\frac{\hat{p_0}}{\phi_{MC}} - 832)$, R=0.93.  

\textbf{Absorption estimation with deep learning} was achieved by training a modified U-Net~\cite{ronneberger2015u} in a supervised manner on the training data. The U-Net was used without a final activation layer, to make it suitable for regression tasks~\cite{grohl2018confidence}. To make use of the digital twin simulations during training, we defined the U-Net to have a two-channel output: the first channel to estimate $\mu_a$ and the second channel to estimate $\phi$. Our U-Net implementation used strided convolutions with a (3$\times$3) kernel size for downsampling and strided transpose convolutions with a (2$\times$2) kernel size for upsampling. It used leaky rectified linear units as the non-linear activation layers and has four sampling steps with a total of 1024 channels in the bottleneck and used the mean squared error as the loss function for both output channels. 

We trained two U-Nets: one on the simulated dataset (DL-Sim, Fig.~\ref{fig:methods2}C) and one on the experimentally acquired training data set (DL-Exp, Fig.~\ref{fig:methods2}D). The training data were normalised by the mean and standard deviation of the respective training data set. The networks were trained for 200 epochs with an initial learning rate of $10^{-4}$ and a learning rate scheduler that reduced the learning rate by 10\% on a plateau of the validation loss for more than 10 epochs. We used an ensemble of five U-Nets trained with five-fold cross-validation where each fold had 16 distinct training phantoms as a validation set.

After training, absorption estimates $\hat{\mu_a}$ were made on unseen experimental PA  images $\hat{p_0}$, after normalising by the mean and standard deviations of the entire experimental training data set ($\mu_{\text{Exp}}$ and $\sigma_{\text{Exp}}$): $
    \hat{\mu_a} = \text{DL}\left(\frac{\hat{p_0} - \mu_{\text{Exp}}}{\sigma_{\text{Exp}}}\right).$

\subsection{Performance evaluation}

For the final performance evaluation, the estimates of each of the five training folds were averaged. In addition to performance evaluation on the held-out test phantoms, we also applied the network to completely different experimental data without additional training. To this end, we used a blood flow phantom experiment and \textit{in vivo} imaging of healthy mice.  While these experiments allow us to investigate the generalisability and limitations of the models, inaccuracies are expected due to variability in the Gr\"uneisen parameter distribution compared to the phantoms.

\textbf{Flow phantom data.} A variable oxygenation blood flow phantom was imaged as described previously~\cite{gehrung2019development}. We fabricated agar-based cylindrical phantoms with a radius of 10\,mm and embedded a polyvinyl chloride tube (i.d. 0.8\,mm, o.d. 0.9\,mm) in the centre of the phantom during the agar setting process at room temperature. For the base mixture, 1.5 \% (w/v) agar (Sigma Aldrich 9012-36-6) was added to Milli-Q water and was heated until it dissolved. Once the mixture had cooled to approximately 40$^\circ$\,C, 2.08\% (v/v) of pre-warmed intralipid (20\% emulsion, Merck, 68890-65-3) and 0.74\% (v/v) Nigrosin solution (0.5 mg/ml in Milli-Q water) was added, mixed, and poured into the mould. PA images of the phantom were taken between 700\,nm and 900\,nm in 20\,nm steps.

\textbf{In vivo mouse data.} All animal procedures were conducted in accordance with project and personal licences (PPL number: PE12C2B96), issued under the United Kingdom Animals (Scientific Procedures) Act, 1986 and approved locally under compliance form CFSB2317. Nine healthy 9-week-old female C57BL/6 albino mice were imaged according to a previously established standard operating procedure \cite{joseph2017evaluation}. Imaging was performed at 10 wavelengths between 700 and 900\,nm averaging over 10 scans each. The kidneys, spleen, spine, and aorta were segmented in PA images using MITK.

\textbf{Performance measures.} In this work, we report the relative error, which is defined as $\text{Rel. Error}(x, \hat{x}) = \frac{|x - \hat{x}|}{x} * 100$, and the absolute error, which is defined as $
\text{Abs. Error}(x, \hat{x}) = |x - \hat{x}|$, where $\hat{x}$ is the estimate for a quantity of interest $x$.  We compute the mean of the pixel-wise measure for each segmentation region. When reporting aggregate results, we proceed to compute the median and interquartile range of the resulting distribution of mean errors per segmentation region. We also report the generalised contrast-to-noise ratio, which is defined as $ \text{gCNR} = 1 - \sum_{k=0}^{N-1}\text{min}(h_i(k), h_b(k))$, where $h_i$ and $h_b$ are the histograms of the signals in the inclusion and background and $k$ is the index of the histogram bin~\cite{rodriguez2019generalized}. The average in a certain \textit{background} segmentation class was taken as the mean, while the average in a \textit{inclusion} class was the median over the first 1.28\,mm depth inside the inclusions. The depth threshold was manually determined sweeping from 0\,mm to 5\,mm depth to yield the best $\mu_a$ quantification results on the training data when using the GT-$\phi$ model. GT-$\phi$ is the model expected to be most affected by the depth-dependent signal decrease.

\subsection{Statistics}
The degree of linear correlation between variables is analysed using the Pearson correlation coefficient. Statistical significance is tested using a non-parametric unpaired Mann-Whitney U test. The significance is indicated by * ($P<0.05$), ** ($P<0.01$), *** ($P<0.001$) and **** ($P<0.0001$). Data are reported as mean $\pm$ standard deviation unless otherwise stated.

\section{Results}
\label{sec:results}

\subsection{Deep learning models show improved spatial distribution of absorption estimates}

We first inspected the images of estimated $\mu_a$ over all test phantoms. We highlight here a particular example of a challenging test case with a halo artefact around one inclusion where DL-Exp achieved a good agreement with the reference in the background (DL-Exp: $0.12\pm 0.07$ cm$^{-1}$ vs Reference: $0.098$ cm$^{-1}$) and one inclusion (DL-Exp: $1.0\pm 0.12$ cm$^{-1}$ vs Reference: $1.04$ cm$^{-1}$) but showed poorer performance in the other inclusion (DL-Exp: $1.97\pm 0.35$ cm$^{-1}$ vs Reference: $3.1$ cm$^{-1}$) at 800\,nm, illustrated using a line profile through both inclusions (Fig.~\ref{fig:results3}). More examples showing success and failure cases are illustrated in Supplementary Figs 4 and 5 for a range of inclusion types and positions. The na\"ive linear calibration model systematically underestimated $\mu_a$ in both inclusions and in the background material (Cal., Fig.~\ref{fig:results3}A). Fluence correction with Monte Carlo simulations yields an overestimation of $\mu_a$ for the larger inclusion and towards the centre of the background material, creating distorted edges of the inclusions (GT-$\phi$, Fig.~\ref{fig:results3}B). 

\begin{figure}[h!tb]
    \centering
    \includegraphics[width=\columnwidth]{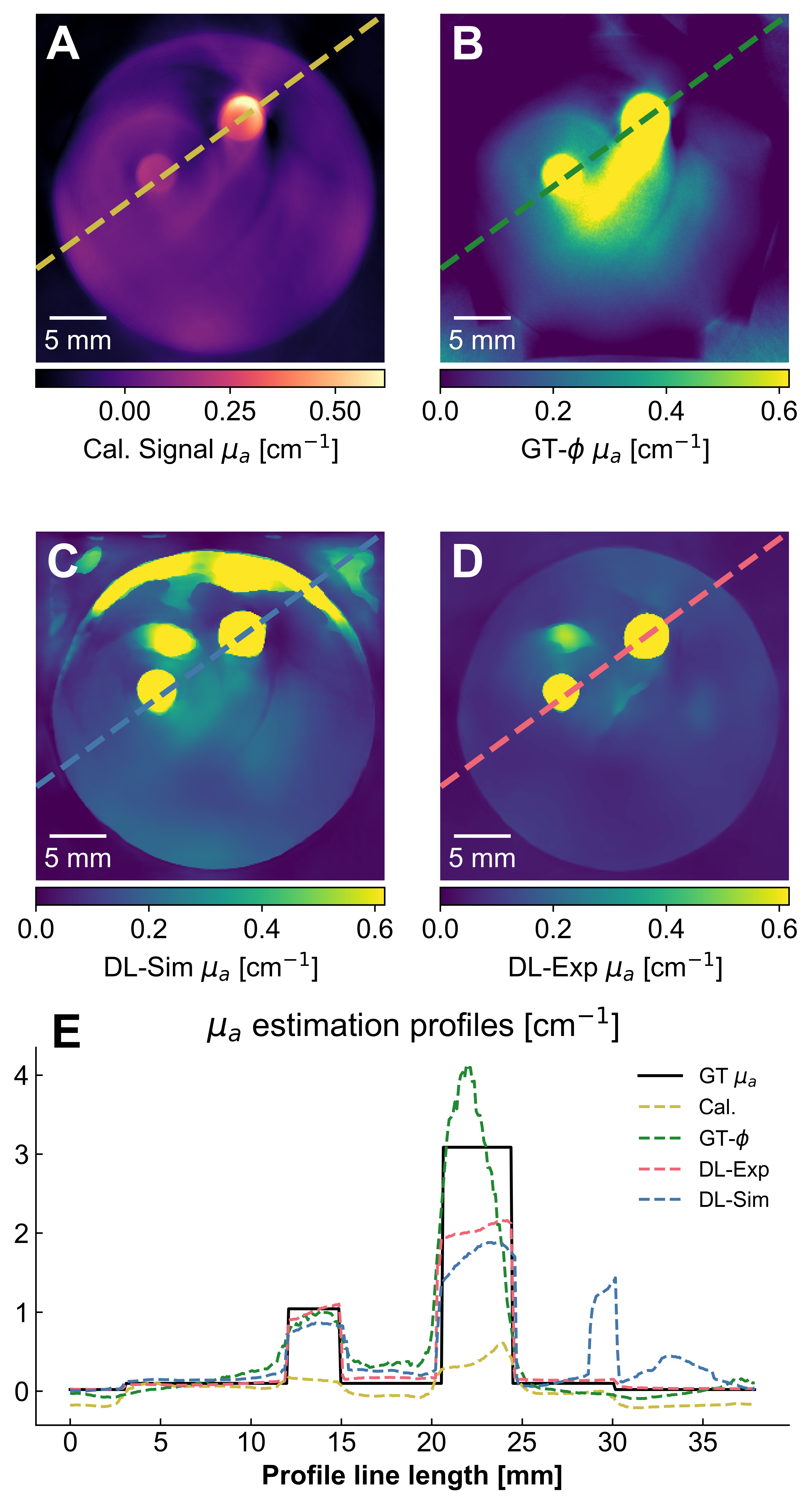}
    \caption{\textbf{Representative performance example from the held-out test set.} (\textbf{A}) Baseline model (Cal., yellow); (\textbf{B}) Fluence correction with reference simulations (GT-$\phi$, green); (\textbf{C}) U-Net trained on simulated data (DL-Sim, blue); and (\textbf{D}) U-Net trained on experimental data (DL-Exp, red). DL-Exp achieved a good agreement with the reference in the background (Reference: $0.098$ cm$^{-1}$) and one inclusion (Reference: $1.04$ cm$^{-1}$) but showed poorer performance in the other inclusion (Reference: $3.1$ cm$^{-1}$) at 800\,nm), as shown through line profiles (\textbf{E}) in different colours. The methods are colour coded in their dedicated colours and the ground truth optical absorption is shown in a solid black line (GT $\mu_a$). The level window in (\textbf{A-D}) is adjusted to clearly visualise the water background, the background material, and the inclusions. }
    \label{fig:results3}
\end{figure}

The estimation of $\mu_a$ from DL-Sim shows a systematic overestimation of the $\mu_a$ in the background material at the upper rim of the phantom, presumably caused by incorrect assumptions present in the 2D acoustic forward model (DL-Sim, Fig.~\ref{fig:results3}C). DL-Sim also shows artefacts, with estimation of a high $\mu_a$ outside of the phantom area in the coupling medium. In contrast, DL-Exp predicts spatially smooth estimates with sharp borders around the phantom and inclusions (DL-Exp, Fig.~\ref{fig:results3}D). DL-Exp also shows improved quantitative performance in the inclusions compared to DL-Sim, particularly on the right inclusion (Fig.~\ref{fig:results3}E). The same findings are seen consistently in line profiles across the test phantom set (see Supplementary Figs 4 and 5), with DL-Exp consistently outperforming all methods in spatial accuracy and quantitative estimation, highlighting the substantial added value of training on experimental data.

\subsection{Absorption coefficient estimates at depth can be recovered with deep learning}
\label{subsec:testset}

\begin{figure}[h!tb]
    \centering
    \includegraphics[width=\columnwidth]{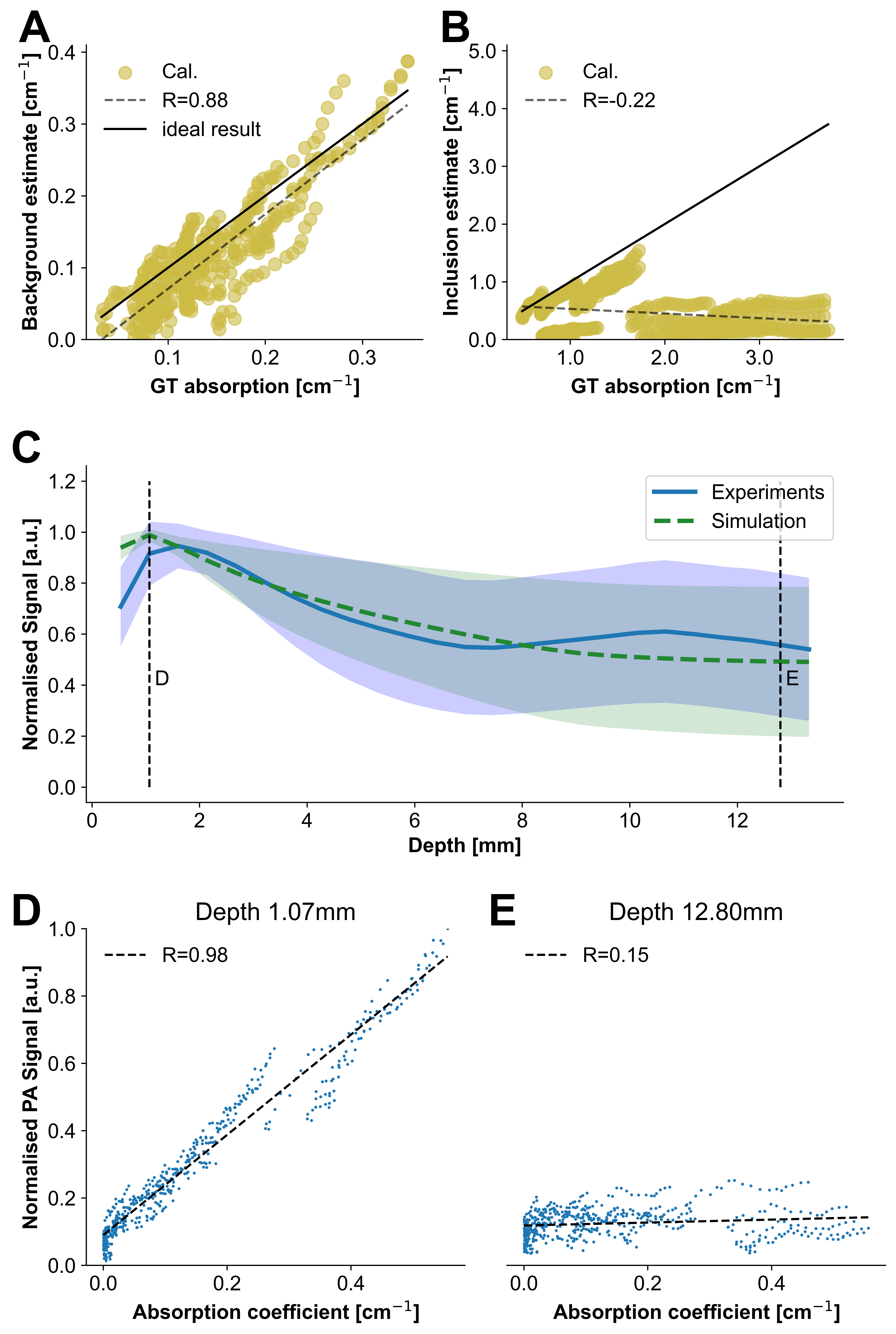}
    \caption{\textbf{PA signal calibration against reference optical absorption with a linear mapping function fails at depth.} Linear mapping derived from the training set was applied to the test set signals from the background (\textbf{A}) and inclusions (\textbf{B}). (\textbf{C}) Experimental and simulated PA signals are shown as function of depth normalised to their maximum signal intensity. The shaded areas correspond to the standard deviation of the signal intensities. Vertical dashed lines highlight depths of approximately 1\,mm and 13\,mm, for which (\textbf{D}) and (\textbf{E}) show the scatter plot and linear fit of all data points across all phantoms in the test set found at those respective depths.}
    \label{fig:results1}
\end{figure}

We next analysed the quantitative performance of the models in more detail. Cal. performs extremely well in the background material at the surface of the phantoms (R=0.88), providing an accurate estimate of $\mu_a$ independently of $\mu_s$ (Fig.~\ref{fig:results1}A). This is not true for signals arising from deeper within the phantoms. In the inclusions at depth within the phantoms, there was no correlation with $\mu_a$ (R=-0.22) (Fig.~\ref{fig:results1}B), as expected due to depth-dependent optical attenuation and spectral colouring~\cite{cox2012quantitative}.

To understand the limit at which the model breaks down, we analysed the 25 training phantoms without any inclusions, showing that the correlation between PAI signal and $\mu_a$ systematically decreases with depth (Fig.~\ref{fig:results1}C), ranging from R=0.98 at 1\,mm depth to R=0.15 at 13\,mm depth (Fig.~\ref{fig:results1}D-E). Although we observe the same overall trend between the experiment and simulation (Fig.~\ref{fig:results1}C) the experimental results do deviate from the monotonic exponential decay seen in the simulation. A number of factors could contribute to this observation: (1) the optical forward model assumes a literature-derived anisotropy and the applicability of the Henyey-Greenstein scattering function~\cite{henyey1941diffuse}, which might differ in our phantom materials; (2) we do not model the noise present in the experimental data; and (3) we do not perform 3D acoustic forward modelling, thus not accounting for out of plane pressure waves or ultrasound focusing.


\begin{table}[h!tb]
\centering
\caption{Comparison of performance of the different methods. Values are shown as median $\pm$ interquartile range/2. R = Pearson correlation coefficient; gCNR = generalised Contrast-to-Noise Ratio}
\label{tab:my_label}
\begin{tabular}{lcccc}
\textbf{Method} & R & gCNR & Rel. Err. & Abs. Err.\\
& [0, 1] & [0, 1] & [\%] & [cm$^{-1}$]\\
\hline
\\
\textbf{Background} \\
\\
Cal. Signal & 0.88 & N/A & 25 $\pm$ 17 & 0.03 $\pm$ 0.02\\
GT-$\phi$ & 0.91 & N/A & 17 $\pm$ 9 & 0.02 $\pm$ 0.02\\
DL-Sim & 0.87 & N/A & 15 $\pm$ 15 & 0.02 $\pm$ 0.02\\
DL-Exp & 0.81 & N/A & 13 $\pm$ 9 & 0.02 $\pm$ 0.01\\
\\
\textbf{All Inclusions} \\
\\
Cal. Signal & -0.22 & 0.73 $\pm$ 0.2& 83 $\pm$ 30 & 1.15 $\pm$ 1.0\\
GT-$\phi$ & 0.84 & 0.95 $\pm$ 0.1 & 22 $\pm$ 14 & 0.30 $\pm$ 0.2\\
DL-Sim & 0.61 & 0.95 $\pm$ 0.1 & 37 $\pm$ 14 & 0.51 $\pm$ 0.5\\
DL-Exp & 0.68 & 0.98 $\pm$ 0.0 & 29 $\pm$ 19 & 0.39 $\pm$ 0.4\\
\\
\textbf{$\mu_a\leq2.5$\,cm$^{-1}$} \\
\\
GT-$\phi$ & 0.83 & 0.86 $\pm$ 0.1 & 21 $\pm$ 13 & 0.23 $\pm$ 0.2\\
DL-Sim & 0.75 & 0.92 $\pm$ 0.1 & 31 $\pm$ 15 & 0.37 $\pm$ 0.2\\
DL-Exp & 0.87 & 0.97 $\pm$ 0.0 & 21 $\pm$ 14 & 0.24 $\pm$ 0.2\\
\\
\end{tabular}
\end{table}

Having established the performance limits of the baseline model (Cal.), we then examined the performance of the U-Net estimations (complete data shown in Supplementary Fig. 6). Application of these maintained or improved the excellent linear mapping in the background material (Table.~\ref{tab:my_label} Background), while recovering a linear relationship between the absorption estimate in the inclusions and the ground truth (Table.~\ref{tab:my_label} Inclusions), showing a greater robustness to depth-dependent effects. Correcting the signals with the Monte Carlo fluence reference yielded the best quantitative result with a median error of 22\% across the entire dataset. Unfortunately, it shows a clear SNR-dependency in performance, with poorer performance at depth, and is most challenging to apply in living subjects due to the lack of knowledge of ground truth optical properties. 

While both deep learning methods underperform compared to GT-$\phi$ (vs. DL-Exp p=0.012, vs. DL-Sim p=10$^{-13}$), DL-Exp (median error of 29\%) significantly (p=10$^{-7}$) outperformed DL-Sim (median error of 37\%). We also observed a systematic underestimation of high absorption coefficients for both deep learning models. Since this effect is consistent for the simulated and experimental data set, we assume that this effect is caused by the composition of the dataset and a larger, more diverse data set would be required to counteract this effect. When evaluating inclusions with $\mu_a\leq2.5$\,cm$^{-1}$, a likely \textit{in vivo} scenario, we find that DL-Exp begins to outperform GT-$\phi$ with regard to the correlation coefficient and gCNR (Table.~\ref{tab:my_label} $\mu_a\leq2.5$\,cm$^{-1}$), demonstrating the future potential of the approach in living subjects.

\subsection{Experimentally trained deep learning models are transferable to a blood flow phantom set}
\label{subsec:flowphantom}

We next applied the Cal., DL-Sim, and DL-Exp $\mu_a$ estimation methods in an independent flow phantom experiment (Fig.~\ref{fig:results6}); the GT-$\phi$ method could not be applied as ground truth optical properties are unknown. 

\begin{figure}[h!tb]
    \centering
    \includegraphics[width=\columnwidth]{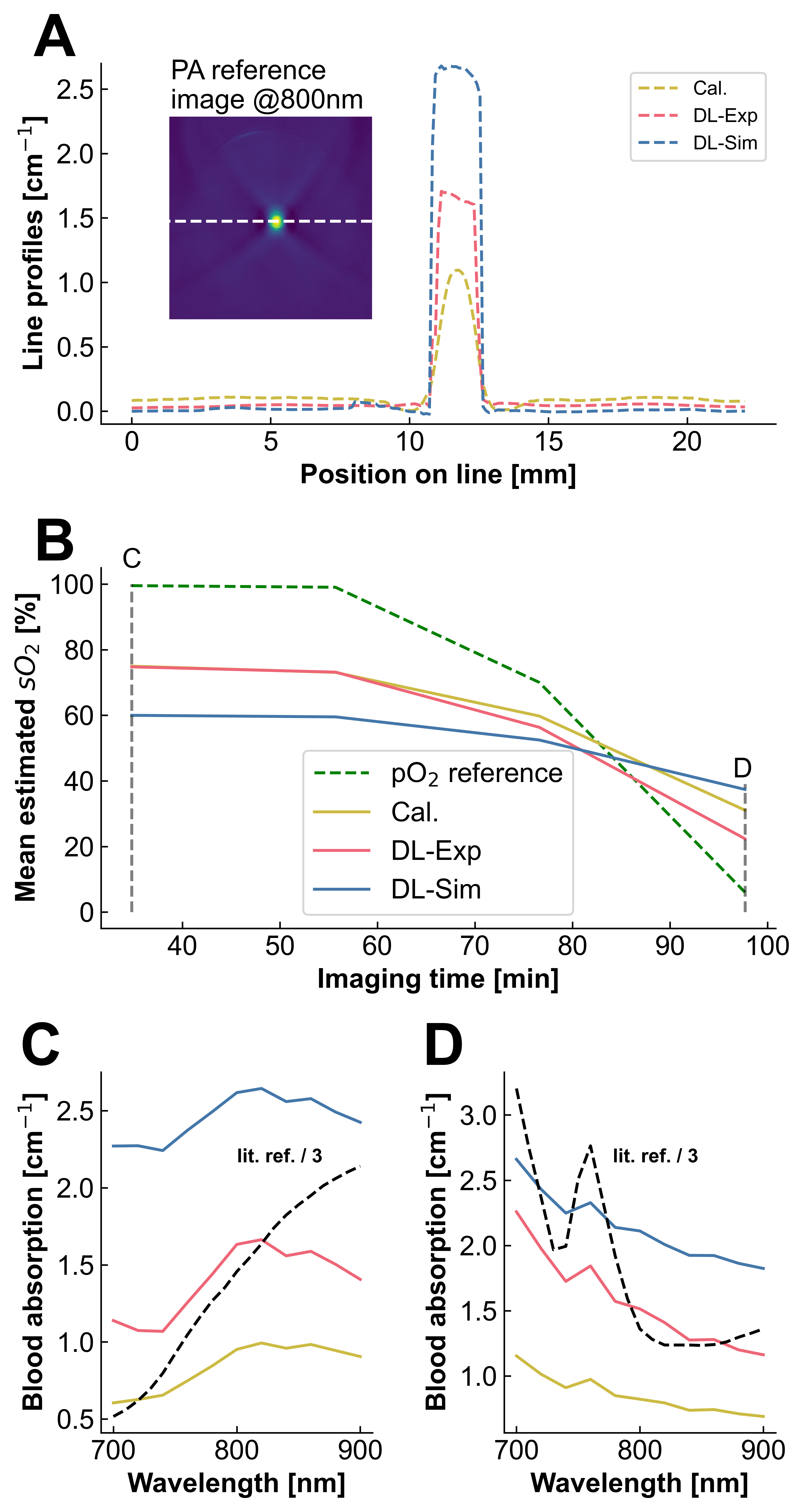}
    \caption{\textbf{DL-Sim fails to preserve the dynamic range of oxygenation response.} \textbf{A} Cross-sectional slice through the phantom with the blood-carrying tube and the spatial absorption profiles along the dashed line. \textbf{B} Mean blood oxygen saturation (sO$_2$) derived with linear unmixing from multispectral $\mu_a$ estimates for each of the methods over time, with spectra at specific time points shown respectively in \textbf{C} and \textbf{D}. A reference for the expected $\mu_a$ spectrum of blood (scaled down by a factor of 3 for visualisation purposes) at the time points is shown as a dashed black line.}
    \label{fig:results5}
\end{figure}

The flow phantom was distinct from the training and test material, as it was fabricated from a different base and included a tube carrying blood that was chemically deoxygenated over time. We found that DL-Exp estimated a nearly two-fold higher $\mu_a$ value in the blood tube and DL-Sim a nearly three-fold increased $\mu_a$ value compared to the baseline method (Fig.~\ref{fig:results6}A). Thus, both methods were able to recover a value closer to the expected reference of about 4.3\,cm$^{-1}$ at 800\,nm~\cite{jacques2013optical}. When computing sO$_2$ from the $\mu_a$ estimates, DL-Sim produced a very poor dynamic range of sO$_2$ during the imaging time, ranging from 60\% to 45\%, rendering any biological applications of the derived biomarker potentially useless (Fig.~\ref{fig:results6}B). DL-Exp, on the other hand, was able to reproduce sO$_2$ at the start of the experiment and marginally increased the dynamic range towards the end of the experiment compared to the reference Cal. method.

Comparison of the mean spectra at different time points (Fig.~\ref{fig:results6}C,D) with literature-derived reference spectra for the $\mu_a$ of haemoglobin~\cite{jacques2013optical}, however, revealed that none of the methods were able to compensate for the spectral colouring induced by water absorption in the coupling media, clearly visible from 850-900\,nm.

\subsection{Oxygenation maps derived from learned $\mu_a$ estimates show more realistic arterial oxygen saturation}
\label{subsec:mousedata}

\begin{figure}[h!tb]
    \centering
    \includegraphics[width=\columnwidth]{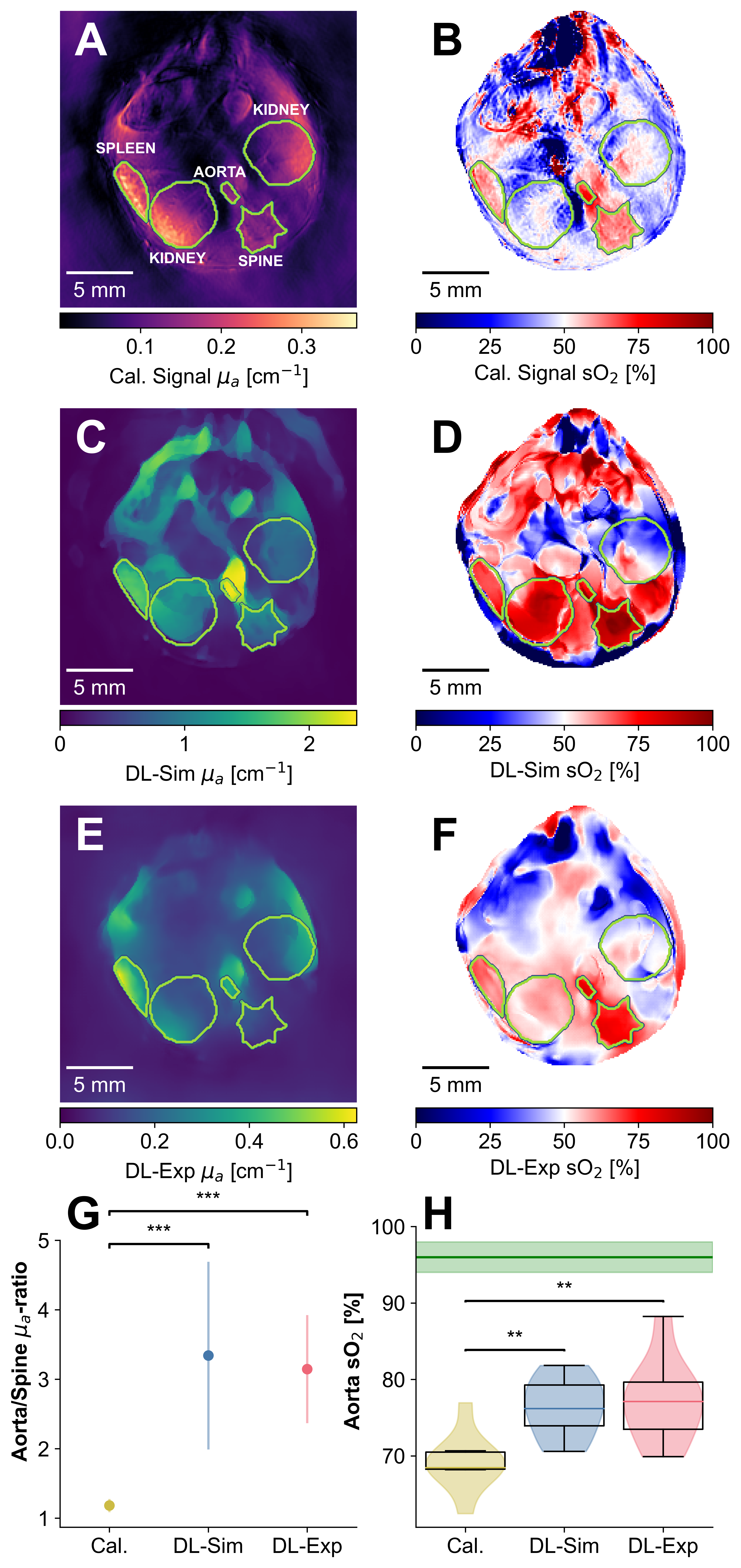}
    \caption{\textbf{Absorption estimation methods applied in living subjects.} (\textbf{A}) Calibrated PA signal (Cal.) and (\textbf{B}) corresponding oxygenation (sO$_2$). \textbf{C)} $\mu_a$ estimates of DL-Sim and (\textbf{D}) corresponding sO$_2$. \textbf{E)} $\mu_a$ estimates of DL-Exp and (\textbf{F}) corresponding sO$_2$. \textbf{G} shows the relative estimated $\mu_a$ ratios between the aorta and spinal region over nine subjects and \textbf{H} shows the distribution of sO$_2$ values in the aorta. The expected sO$_2$ of 96\%$\pm$2\% is shown as a green area.}
    \label{fig:results6}
\end{figure}

Finally, we tasked Cal., DL-Sim, and DL-Exp methods to estimate $\mu_a$ \emph{in vivo} (Fig.~\ref{fig:results6}A,C,E) and subsequently calculated sO$_2$ from multispectral estimates (Fig.~\ref{fig:results6}B,D,F). Despite their excellent spatial performance in phantoms, the U-Net $\mu_a$ estimates were not able to retain the high spatial frequencies of the \textit{in vivo} PA image, which also translates to the resulting sO$_2$ estimates (Fig.~\ref{fig:results6}C,E), most likely due to the lack of high spatial frequency features in the training set. 

We computed the relative $\mu_a$-ratio between the aorta and the spine (Fig.~\ref{fig:results6}G), to understand whether the methods could compensate depth-dependent decreases in signal intensity. Physiologically, we would expect a higher $\mu_a$ and oxygenation in the aorta compared to the spine due to higher average blood volume and presence of both arteries and veins in the spine. While the Cal. estimate gave approximately the same $\mu_a$ for both regions, the U-Net estimates result in a 2-3 fold higher $\mu_a$ in the aorta compared to the spine. Consistent with the artefacts observed in the test phantoms, DL-Sim estimated high $\mu_a$ values in the upper rim of the mouse body ans in the water background.

Application of linear unmixing revealed that the general trends of high and low oxygenation in the segmented organs were quite consistent across all methods. A notable exception was the left kidney, where both U-Net estimations showed relatively high oxygenation, whereas the calibrated signal estimated relatively low oxygenation. Computing the arterial sO$_2$ in the aorta reveals an 8 $\pm$ 5 percentage points increase (p=0.002) in blood oxygenation on the DL-Exp $\mu_a$ estimate compared to the calibrated signal (Fig.~\ref{fig:results6}H), though all models still underestimate the expected arterial blood oxygenation in an anaesthetised mouse of about 94\%-98\%~\cite{loeven2018arterial}, perhaps relating to the generally lower dynamic range seen in the blood phantom.

\section{Discussion}
\label{sec:discussion}
%
We developed a supervised deep learning method to address the challenge of estimating optical absorption coefficients in PAI enabled by a collection of carefully annotated phantoms. The dataset consists of an unprecedented 102 tissue-mimicking phantoms spanning an important biological range of tissue optical properties. When assuming independence of images at 21 captured wavelengths from 700\,nm to 900\,nm, our dataset features a total of 2142 unique PA images of targets with distinct and known $\mu_a$ and $\mu_s'$ distributions. Each image was subject to manual segmentation and quality assessment, then assigned reference optical absorption and scattering coefficients as determined by independent DIS measurements. We used digital twins of the phantoms to simulate a synthetic replica data set to analyse the differences between using simulated and synthetic data sets for model training.\\

\textbf{Discussion of Results.} We show that erroneous model assumptions have an influence on the quantitative accuracy of the deep learning model and that training with experimentally acquired phantoms has advantages in both accuracy and generalisability of the approach. Our results indicate that tissue-mimicking phantoms can be used as reliable calibration objects and that using experimental training data is beneficial compared to synthetic training data, indicating the existence of a simulation gap. At the same time, we confirm that linear calibration schemes fail as they are unable to account for the exponential decay of fluence with depth. We find that using the experimental data set yields improved accuracy across multiple test cases compared to simulated data sets, at least given the simulation pipeline used in this work.


The deep learning models were initially applied to a test phantom set and results were compared with the na\"ive calibration model and fluence compensation using Monte Carlo modelling (assuming known optical properties). All methods could recover some correlation of the estimated $\mu_a$ coefficients compared to the reference $\mu_a$. A striking result of DL-Exp is the overall improvement in spatial accuracy of the predictions, which is demonstrated by smooth estimates with sharply defined borders between segmentation classes, quantified by consistently high gCNR values. On the other hand, we found a median relative quantification error of $\approx 29\%$ when training on experimentally acquired data and $\approx 37\%$ when training on simulated data. In comparison, fluence correction with a Monte Carlo model of the ground truth absorption and scattering properties, achieves a relative absorption estimation error of $\approx 22\%$, but is less accurate in terms of spatial estimation at depth and less generalisable to other settings.

When evaluating using only imaging targets with $\mu_a\leq2.5$\,cm$^{-1}$, we find that DL-Exp begins to outperform GT-$\phi$ on several reported measures. By extending the training data set, we hope to achieve significantly better quantification results. Improvement of the deep learning model to mimic traditional iterative schemes might also be a promising way forward to further improve quantitative performance.


When moving to completely different phantom and experimental data, the U-Nets performed less well in terms of spatial predictions. For example, the $\mu_a$ estimate of blood in the flow phantom was significantly increased although it was still more than a factor of two too low compared to the expected $\mu_a$ from literature. We did not include the Gr\"uneisen parameter or 3D acoustic forward modelling in the digital data twins. Hence, the U-Nets are expected to be sensitive to changes in the underlying phantom material and the simulated images are not as accurate as they could be, which could be major reasons for the remaining observed discrepancies on the flow phantom data set. Similarly for the \textit{in vivo} application in a mouse, DL-Exp gave a much more realistic representation of the absorption difference between the aorta and spine, and resulted in a substantial elevation of the estimated arterial oxygen saturation. The final results remained an under-estimate compared to the expected $\geq 94\%$ but suggest a greater resilience to depth-dependent attenuation.

The U-Nets produced estimates with sharp edges on the blood-background interface in the flow phantom, likely because of the consistent nature of the spatial frequency content of the images. Contrary to the promising spectral performance, when applied to the mouse data, the U-Nets were not able to recover the high spatial frequency content of the image, suggesting a need to enrich the training data set with inclusions that better reflect the complexity of the \textit{in vivo} imaging application in order to achieve high spatial accuracy in the general case.\\


\textbf{Experimental limitations.} While training on experimental phantom measurements shows promise to advance quantitative $\mu_a$ estimates in PAI, there are also limitations that must be addressed in future work. Considering first experimental limitations, we only use titanium dioxide to tune $\mu_s'$ and nigrosin to tune $\mu_a$ in the phantoms. This is important to minimise the influence of the molecule- and concentration-dependent Gr\"uneisen parameter but might introduce a wavelength-dependent bias to the training data that may challenge our assumption of independence of images across wavelengths. We assumed images from different wavelengths to be independent, to maximise the number of training images even though images at neighbouring wavelengths will have a high correlation. While such an assumption could potentially lead to the introduction of a wavelength-dependent bias, our results do not show an indication for it, suggesting that the assumption -- while not being ideal -- did not systematically affect the learned algorithms (cf. Supplementary Fig. 7). If the assumption does not hold, it could impact the quantification of multispectrally derived biomarkers, such as blood oxygenation sO$_2$.

A second experimental limitation of the study is the assumption that DIS measurements can be taken as a ground truth~\cite{zamora2013double}, where in fact they may introduce substantial uncertainty~\cite{beek1997vitro} for example from modelling errors by \textit{a priori} assumptions for the sphere parameters, sphere corrections, or measurement error for sample thicknesses. Accurate quantification of optical properties is highly challenging, with variation of up to 30\% and 40\% for $\mu_a$ and $\mu_s'$, respectively, when quantifying the same sample with different instruments~\cite{pifferi2005performance}. We examined the influence of our uncertainties and found a standard error of 2.5\% in the thickness measurements, which translates to approx. 5\% error of the optical absorption coefficient and simulated initial pressure distribution (cf. Supplementary Fig. 8). In conjunction with the 10\%-20\% error we determined for the DIS system based on cross-reference with time-resolved near-infrared spectroscopy~\cite{hacker2022thesis}, our reference optical material properties are subject to a standard error in the order of 20\%. In future work, these uncertainties could be incorporated into model training by using Bayesian deep learning methods.\\

\textbf{Computational limitations.} We introduced a number of approximations during the Monte Carlo and k-space simulations, such as 2D acoustic modelling,  the laser source geometry and beam profile, and absolute homogeneity of the phantom materials (i.e. sound speed and density). We used the delay-and-sum algorithm as the acoustic inverse operator, which is known to introduce artefacts to the reconstructed images for our imaging setup~\cite{xu2005universal}. We assume that a combination of these modelling errors in conjunction with the presence of measurement noise deep inside tissue is the reason for the relatively high $\mu_a$ quantification error for GT-$\phi$. The influence of noise is highlighted when visualising the absorption estimated within larger inclusions with high $\mu_a$, leading to an increasing overestimation of $\mu_a$ with depth and a wavelength-dependent performance of GT-$\phi$ (cf. Supplementary Fig. 7).\\


\textbf{Future directions.} When only analysing results for absorbers with $\mu_a\leq2.5$\,cm$^{-1}$, our findings suggest that a learning-based approach has distinct advantages that might allow it to eventually outperform fluence correction. Learning-based models might learn to correct for effects such as measurement noise, out-of-plane signal contributions or signal blurring caused by the point spread function of the measurement system. Unfortunately, our results also show that to achieve this, we require: (1) more and broader training data; (2) highly optimised and accurate optical and 3D acoustic forward models; (3) realistic noise models; as well as (4) model-based acoustic inversion schemes to obtain more faithful reconstructions of the initial pressure distributions. Finally, in order to validate the trained methods, diverse \textit{in vivo} test data sets have to be acquired that are representative of the target applications and annotated with high-quality and high-confidence labels of the underlying physiology. Since ground truth knowledge of the underlying optical properties \textit{in vivo} is rarely available, the validation of learned algorithms remains a major roadblock on the path towards quantitative photoacoustics.

\section{Conclusion}
\label{sec:conclusion}

We introduce a collection of well-characterised phantoms to, for the first time, enable supervised training and evaluation of learned quantitative PAI methods on experimental data. Leveraging the power of deep learning and computational modelling of the forward processes involved in PAI, we demonstrate that our phantoms can be used as reliable calibration objects. Our experiments confirm the limitations of linear calibration schemes in accounting for depth-dependent fluence effects and show that a deep learning method trained on experimentally acquired datasets can provide better  optical property estimation of previously unseen test phantoms compared to simulated data sets. With demonstrations in different test scenarios, including in living subjects, deep learning shows potential for real-world applications. Despite remaining limitations in spectral and spatial accuracy, our work represents an exciting advancement towards the development of accurate and reliable quantitative PAI data analysis.

\section*{Bibliography}
\bibliography{bibliography}

\section*{Disclosures}

J. Gr\"ohl is funded by the Deutsche Forschungsgemeinschaft (DFG, German Research Foundation) under project GR 5824/1.\\

L. Hacker was funded by the MedAccel programme of NPL, financed by the Industrial Strategy Challenge Fund of the Department for Business, Energy and Industrial Strategy.\\

Cancer Research UK funds E.V. Bunce (RadNet Cambridge C17918/A28870), T.R. Else, P.W. Sweeney and S.E. Bohndiek (C9545/A29580).\\

The work is supported by the NVIDIA Academic Hardware Grant Program and utilised two Quadro RTX 8000.\\

For the purpose of open access, the authors have applied a Creative Commons Attribution (CC BY) licence to any \textit{Author Accepted Manuscript} version arising from this submission.\\

The data to reproduce the findings of this study will be made available via {\color{blue}{https://doi.org/10.17863/CAM.96644}} upon publication of this manuscript.\\

The respective code work is available on GitHub at {\color{blue} https://github.com/BohndiekLab/end\_to\_end\_phantom\_QPAT}.

\onecolumn
\setcounter{figure}{0}

\section*{Supplemental Materials:}

\section*{Sample preparation and phantom designs}

This section provides additional details on the experimental setup. In particular, we describe the sample preparation and phantom designs used in our study, which were optimized to mimic the optical and acoustic properties of biological tissue.

\begin{figure}[h!tb]
    \centering
    \includegraphics[width=\textwidth]{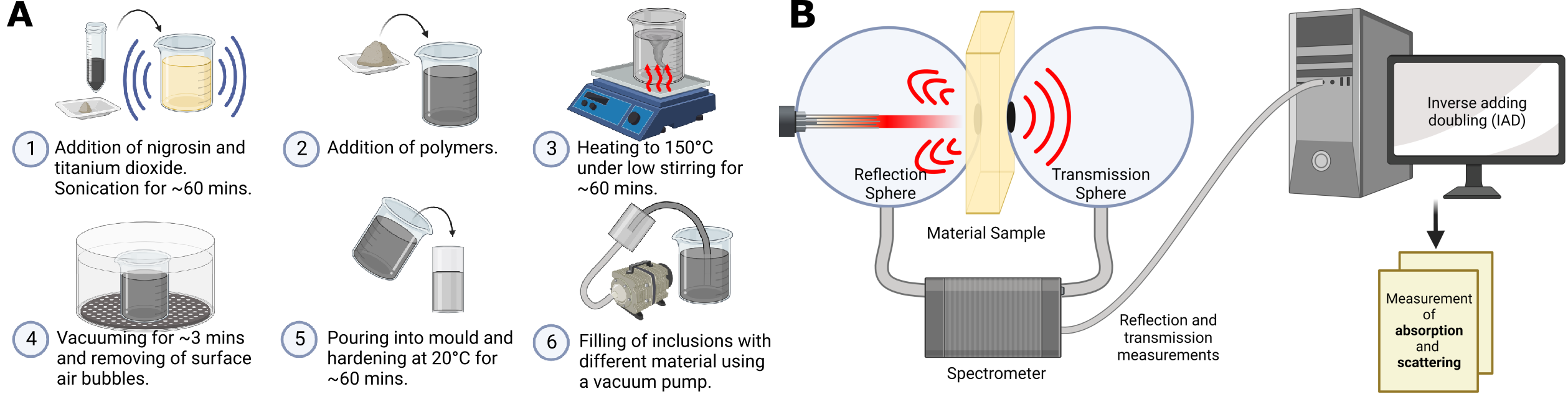}
    \caption{\textbf{Sample preparation and DIS measurements.} \textbf{A} shows the phantom preparation recipe. In brief, the following steps have to be taken: (1) Addition of nigrosin and titanium dioxide and sonication for ~60 mins; (2) addition of polymers; (3) heating to 150$^\circ$ C under low stirring for ~60mins; (4) vacuuming for ~3 mins and removing surface air bubbles; (5) pouring into mould and hardening for ~60 mins at room temperature; and (6) filling of inclusions using a vacuum pump. \textbf{B} A double integrating sphere system is used to measure the total reflectance and diffuse transmittance of the sample. Then, the inverse adding doubling (IAD) algorithm is run. It uses the reflectance and diffuse transmittance measurements to estimate the optical absorption and reduced scattering coefficients of the sample.}
    \label{fig:overview}
\end{figure}

\begin{figure}[h!tb]
    \centering
    \includegraphics[width=\textwidth]{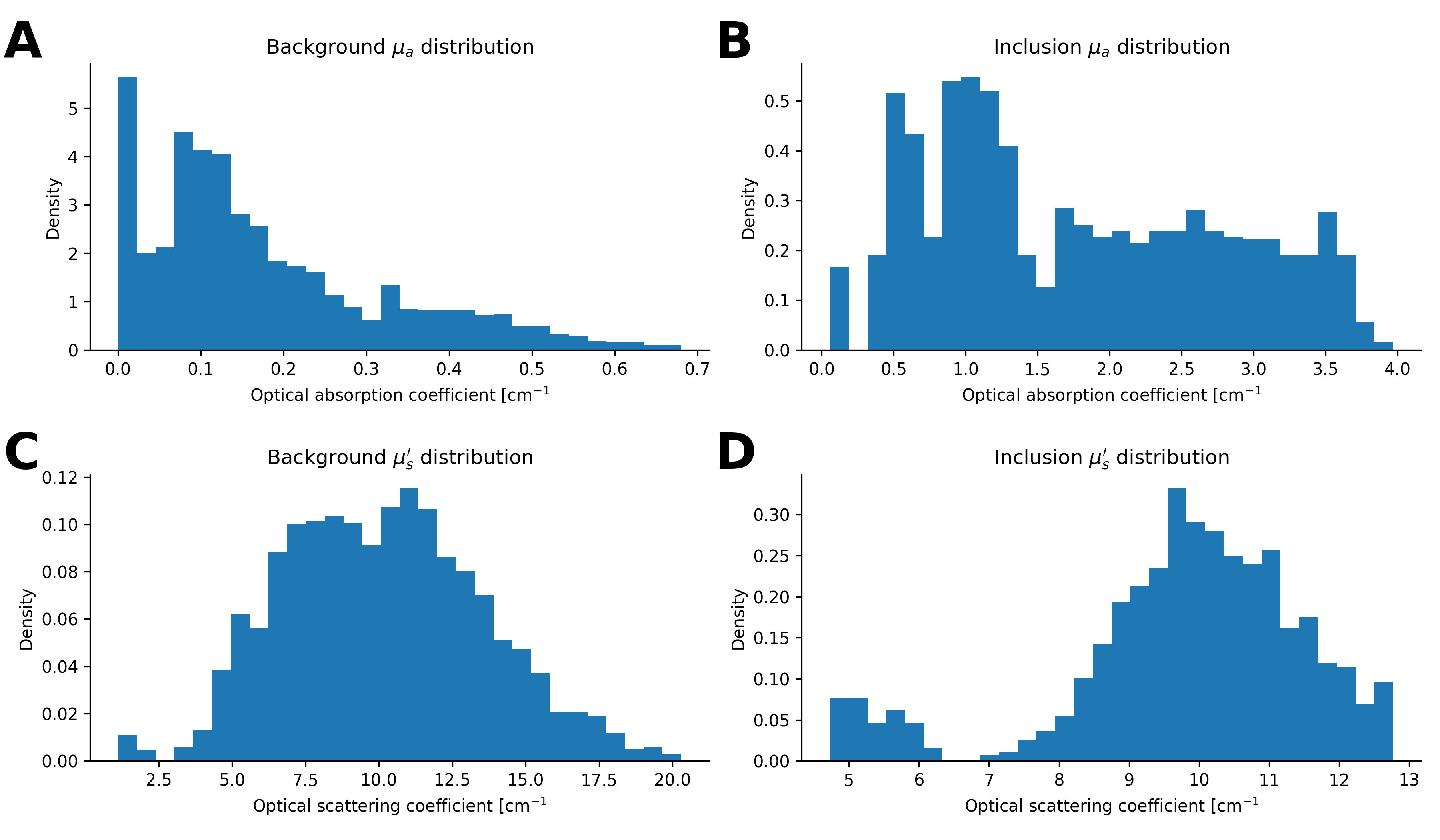}
    \caption{\textbf{Histogram plots showing the distribution of optical properties in a sample.} \textbf{A} Distribution of the optical absorption coefficient in the background material, \textbf{B} distribution of the optical absorption coefficient in the inclusion material, \textbf{C} distribution of the reduced scattering coefficient in the background material, and \textbf{D} distribution of the reduced scattering coefficient in the inclusion material. The histograms provide information on the variability and range of the optical properties in the sample, which is important for understanding the optical behavior of the material and for the development of accurate models and simulations.}
    \label{fig:optical_property_density}
\end{figure}

\clearpage

\section*{Agreement between simulation and experiment}

In this section, we present a comparison between the simulated and experimental measurements, demonstrating a good agreement between the two.

\begin{figure}[h!tb]
    \centering
    \includegraphics[width=0.75\textwidth]{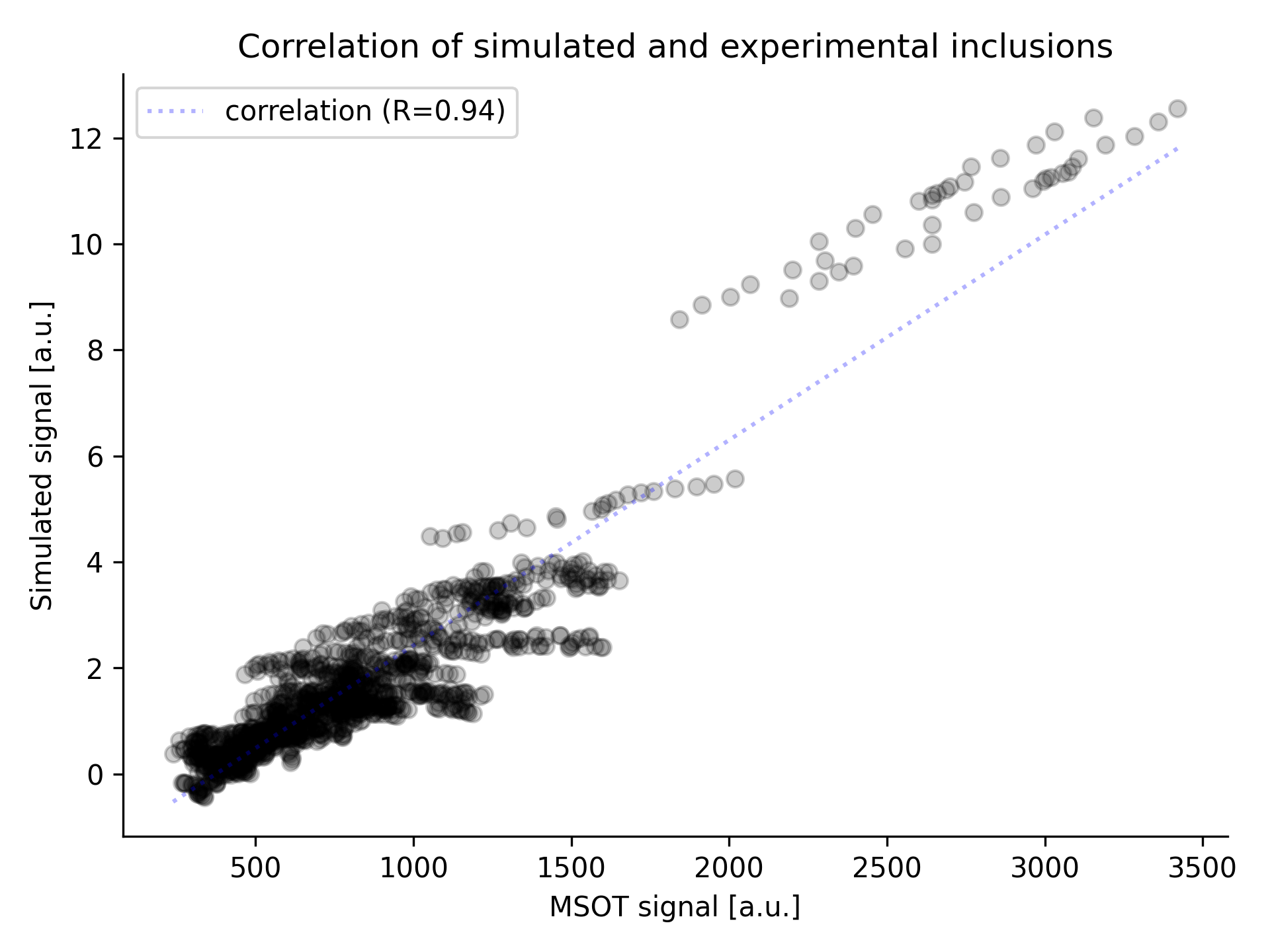}
    \caption{Scatter plot showing the mean experimentally acquired signal in the inclusion targets on the x-axis and the simulated signal amplitude in the same regions on the y-axis. A linear regression of the points shows an excellent Pearson correlation coefficient of 0.94.}
    \label{fig:my_label}
\end{figure}

\clearpage

\section*{Supplemental Result Figures}

This section contains supporting result figures that provide a visual representation of the experimental data, including images of the tissue-mimicking phantoms and PAI results obtained using our deep learning approaches.

\begin{figure}[h!tb]
    \centering
    \includegraphics[width=0.4\textwidth]{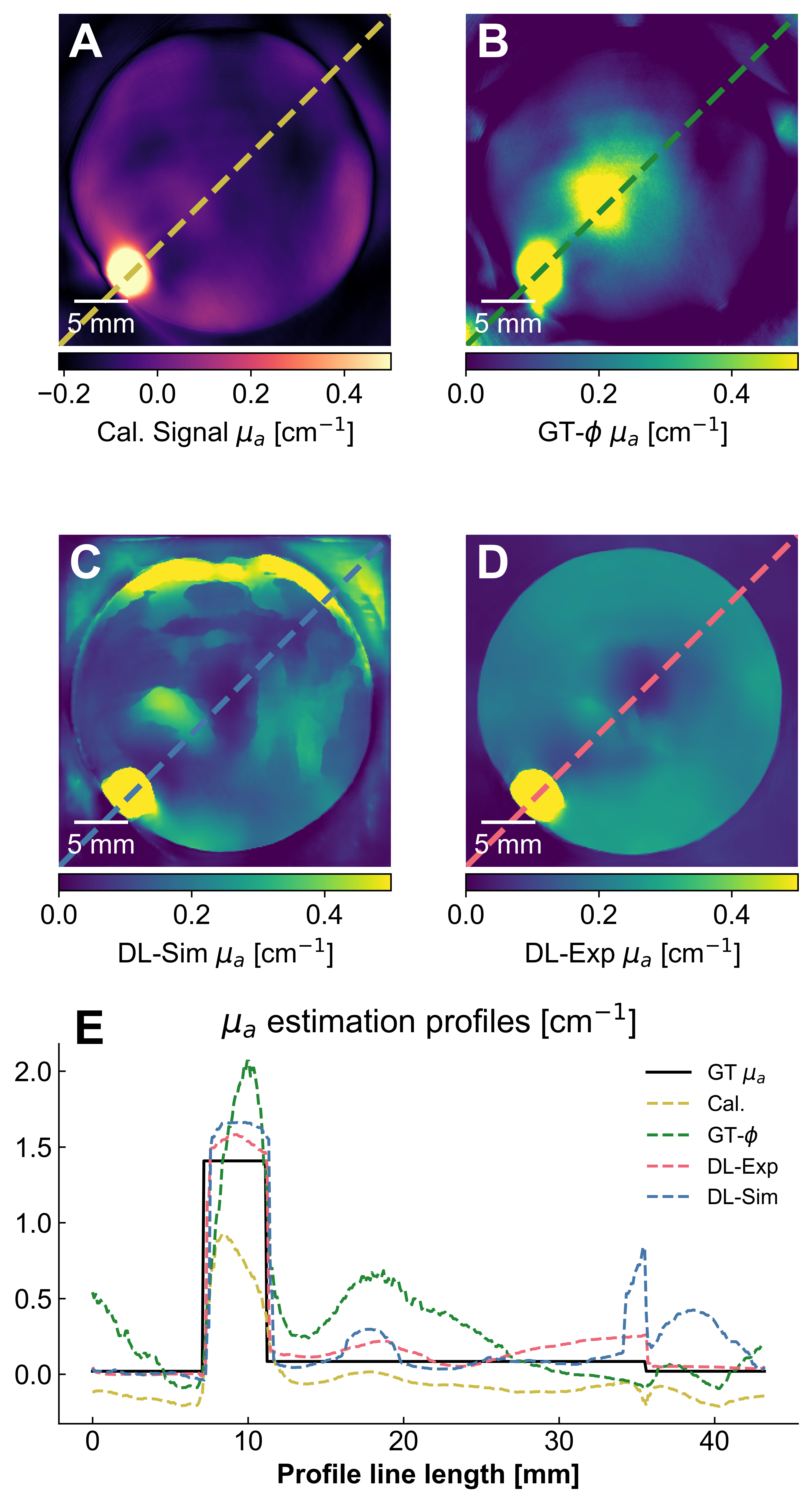}
    \quad\quad\quad\quad\quad\quad\quad\quad
    \includegraphics[width=0.4\textwidth]{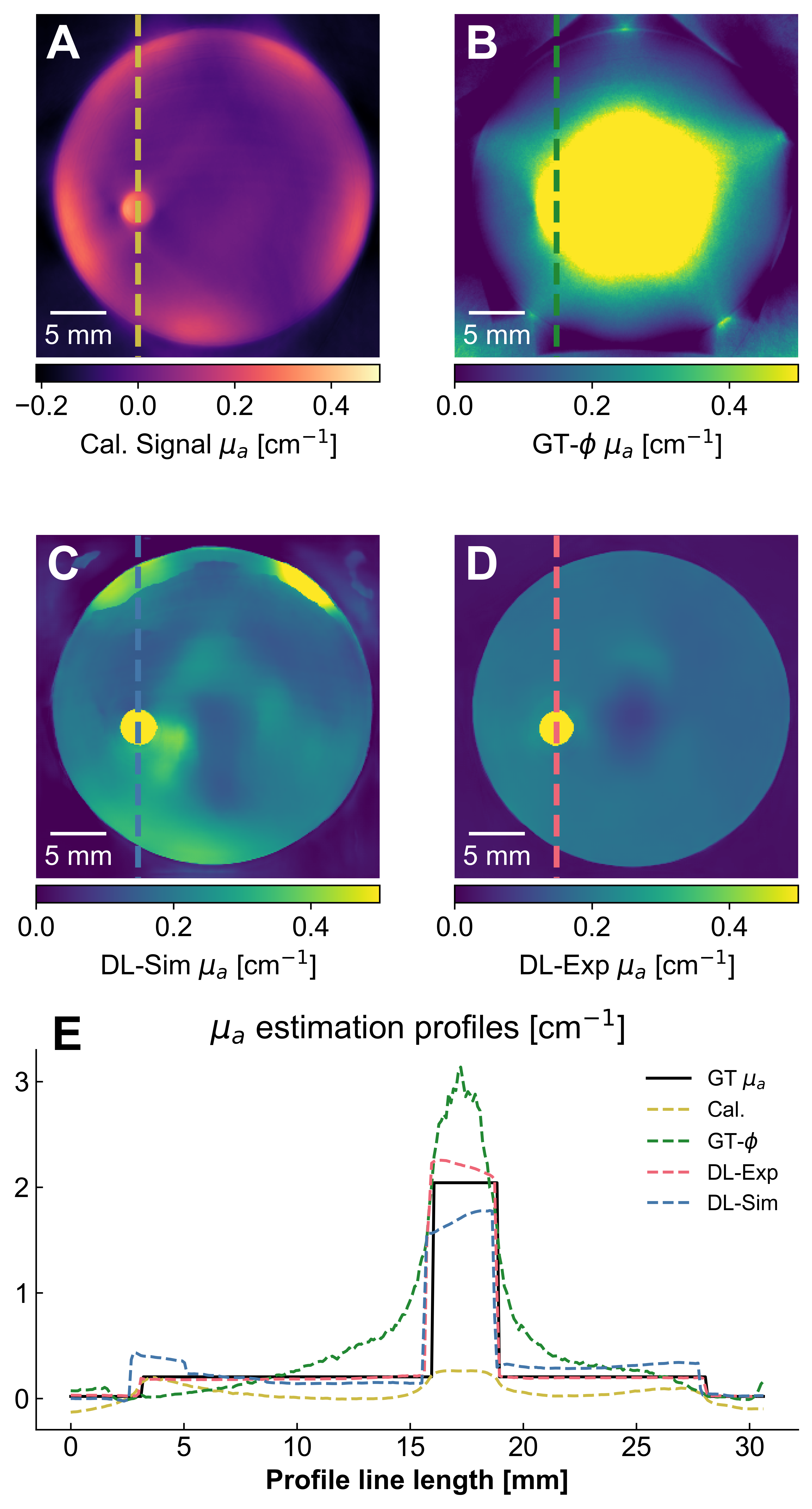}
    \caption{\textbf{Side-by-side plots of two test-set phantoms and the $\mu_a$ estimation results.} \textbf{A} Baseline model (Cal., yellow); \textbf{B} Fluence correction with reference simulations (GT-$\phi$, green); \textbf{C} U-Net trained on experimental data (DL-Exp, red); and \textbf{D} U-Net trained on simulated data (DL-Sim, blue). Line profiles are shown for each method (\textbf{E}) in different colours. The methods are colour coded in their dedicated colours and the ground truth optical absorption is shown in a solid black line (GT $\mu_a$). The level window in \textbf{A-D} are adjusted to clearly visualise the water background, the background material, and the inclusions.}
    \label{fig:example1}
\end{figure}

\begin{figure}[h!tb]
    \centering
    \includegraphics[width=0.4\textwidth]{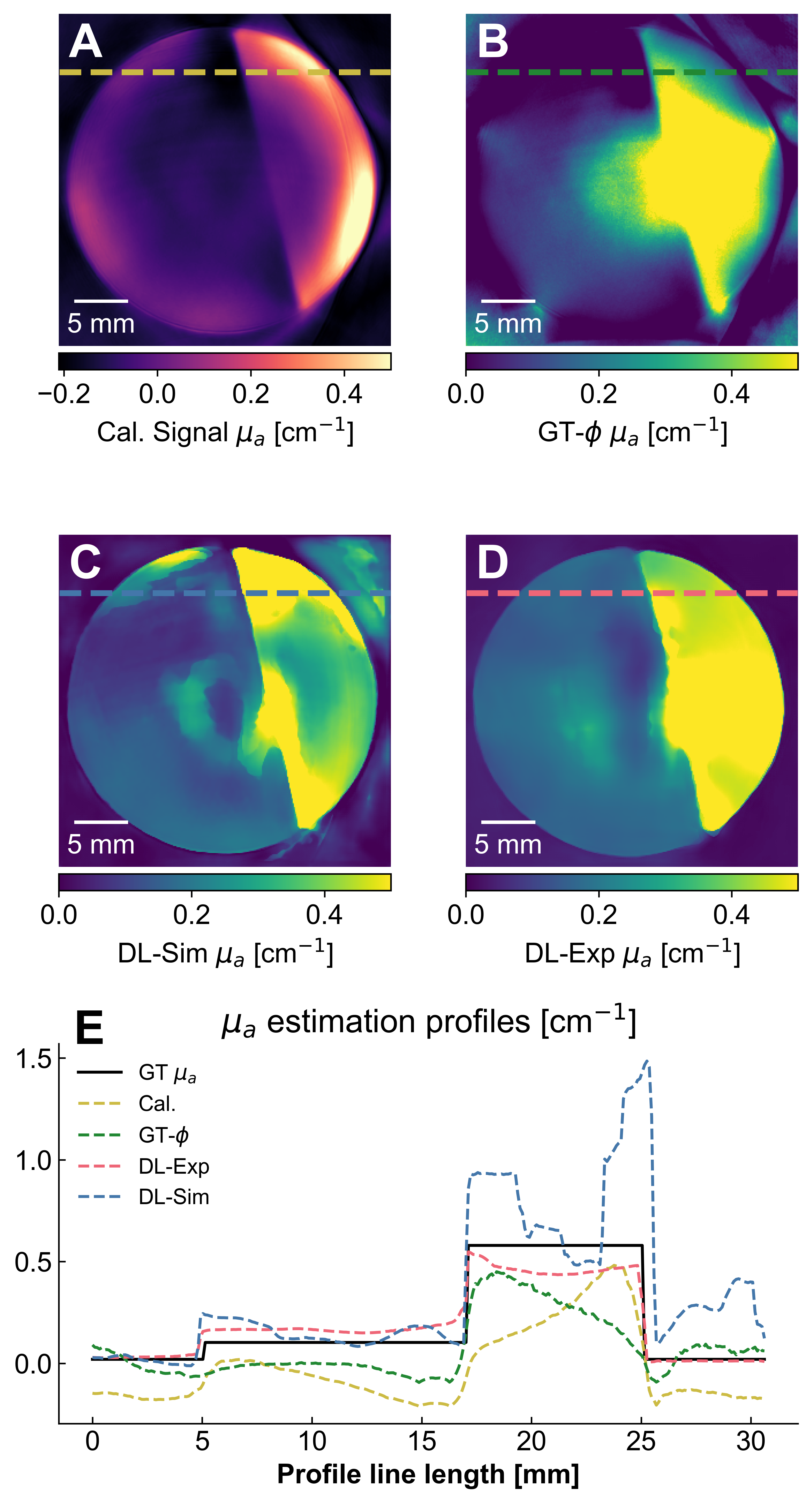}
    \quad\quad\quad\quad\quad\quad\quad\quad
    \includegraphics[width=0.4\textwidth]{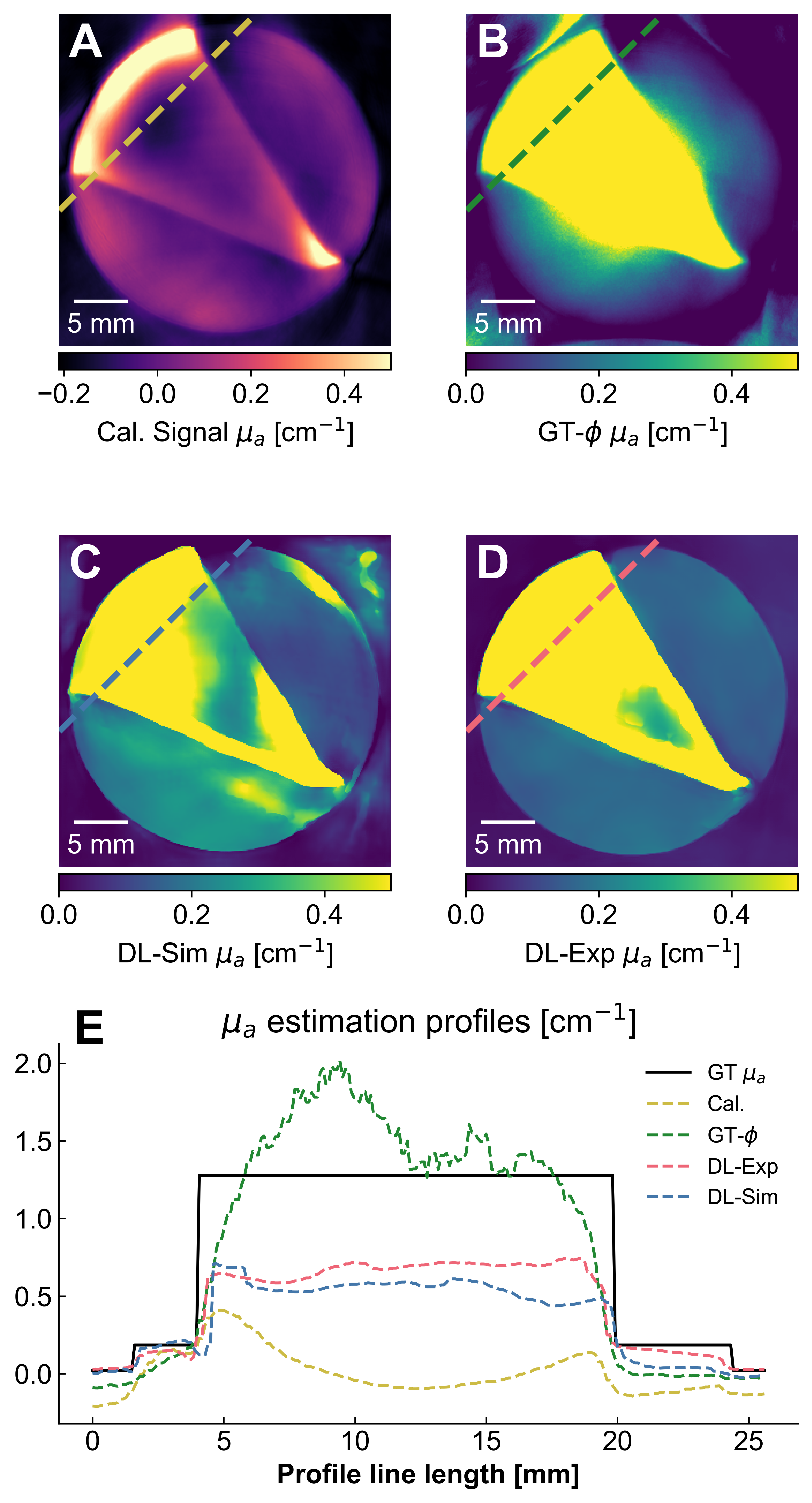}
    \caption{\textbf{Side-by-side plots of two test-set phantoms and the $\mu_a$ estimation results.} \textbf{A} Baseline model (Cal., yellow); \textbf{B} Fluence correction with reference simulations (GT-$\phi$, green); \textbf{C} U-Net trained on experimental data (DL-Exp, red); and \textbf{D} U-Net trained on simulated data (DL-Sim, blue). Line profiles are shown for each method (\textbf{E}) in different colours. The methods are colour coded in their dedicated colours and the ground truth optical absorption is shown in a solid black line (GT $\mu_a$). The level window in \textbf{A-D} are adjusted to clearly visualise the water background, the background material, and the inclusions.}
    \label{fig:example3}
\end{figure}

\clearpage

\begin{figure}[h!tb]
    \centering
    \includegraphics[width=\textwidth]{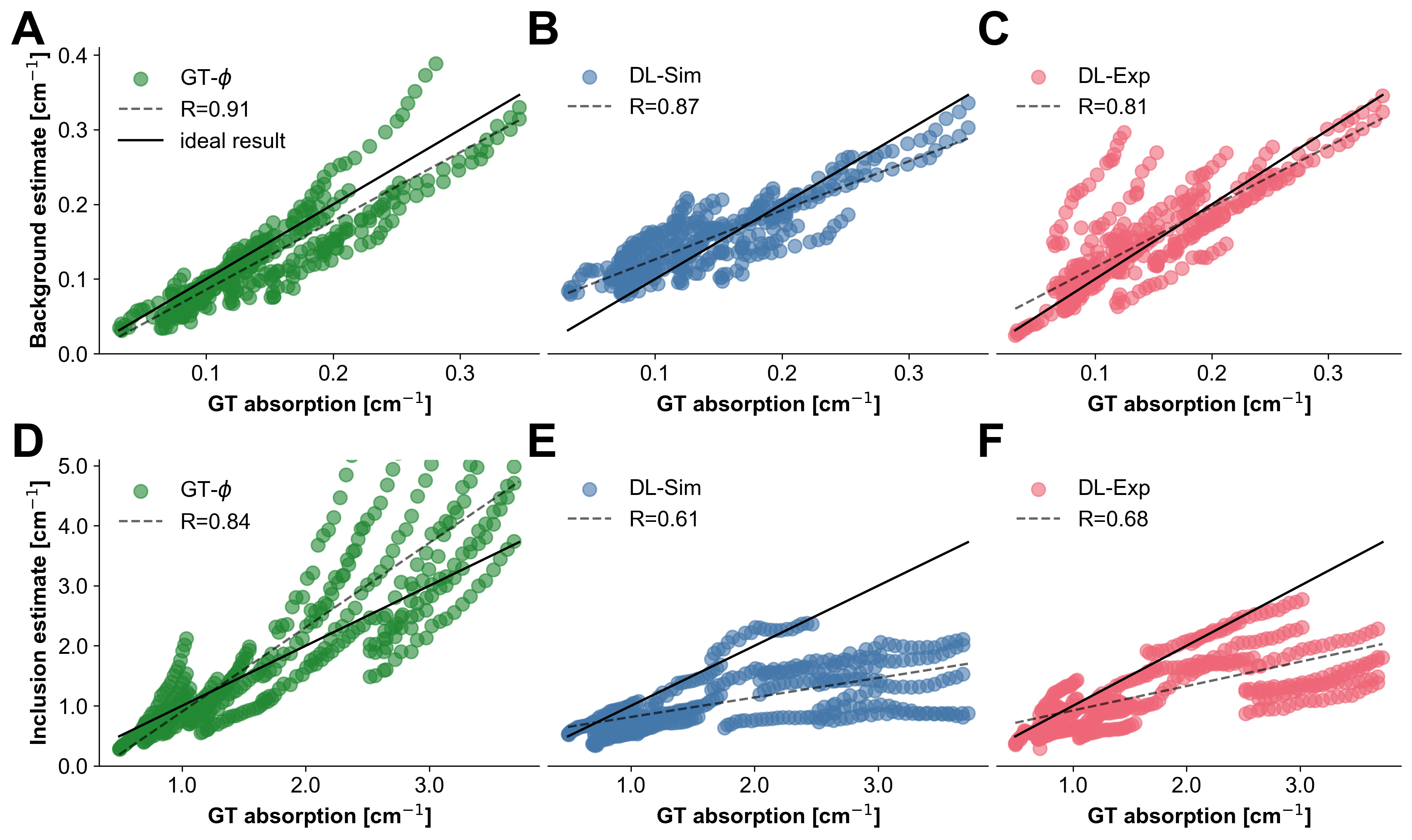}
    \caption{\textbf{Absorption coefficient estimates of the different methods.} \textbf{(A, D)} show the results when correcting with the fluence derived from Monte Carlo simulations (green); \textbf{(B, E)} show the results when estimating with the U-Net trained on simulations (blue); and \textbf{(C, F)} show the results with a U-Net trained on experimentally acquired data (red). \textbf{(A, B, C)} show the results only considering the phantom base material and \textbf{(D, E, F)} show the results only considering the inclusions.}
    \label{fig:my_label}
\end{figure}

\begin{figure}[h!tb]
    \centering
    \includegraphics[width=\textwidth]{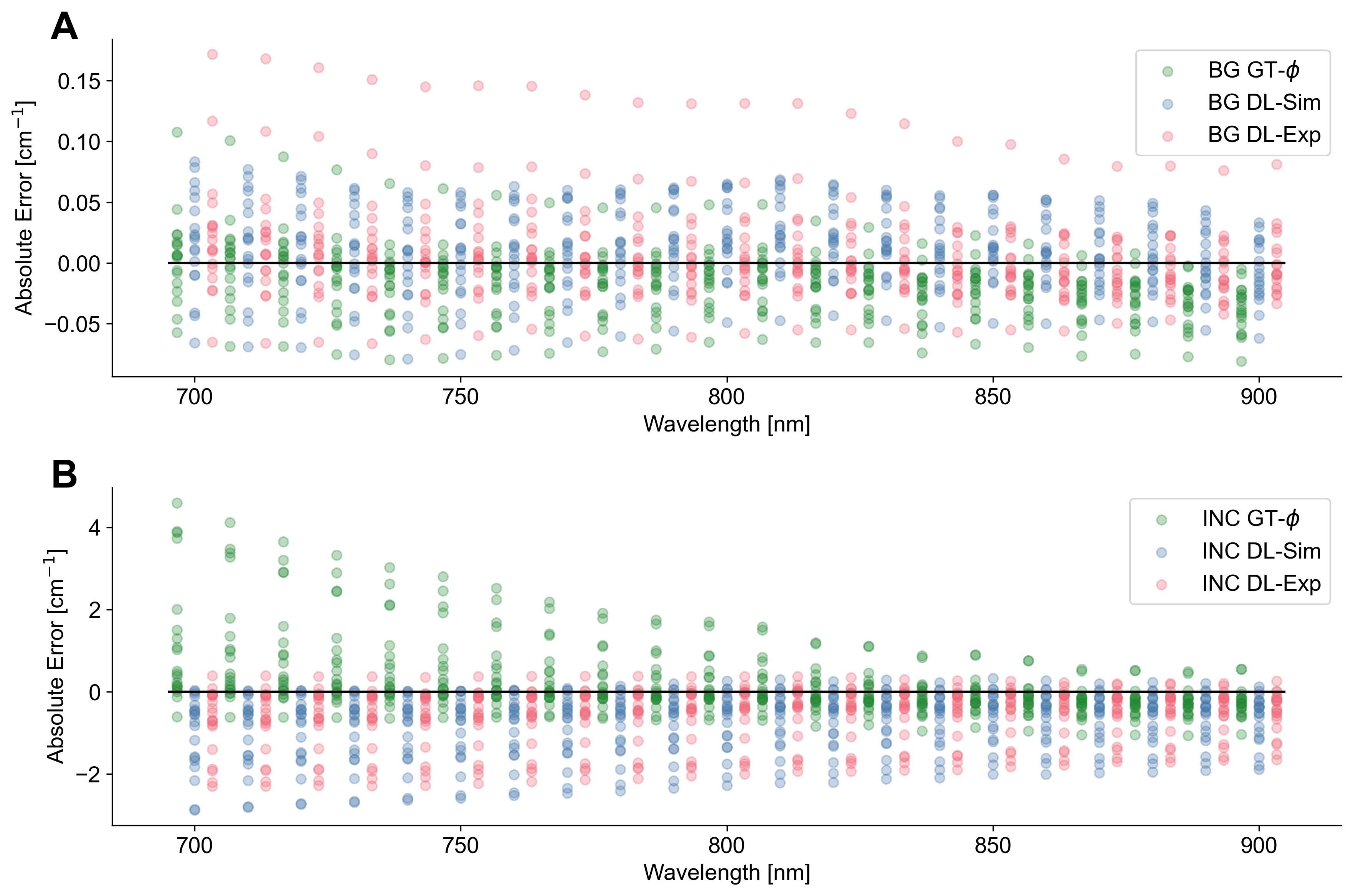}
    \caption{\textbf{The $\mu_a$ estimation errors as a function of wavelength.} \textbf{(A)} shows that no significant bias exists in the background estimates. \textbf{B} reveals that GT-$\phi$ exhibits a monotonously decreasing error with wavelength, caused by the increasing SNR in depth with increasing wavelength. While also exhibiting a slight bias, the wavelength-dependent change in error is much less pronounced for the estimated of the learned methods.}
    \label{fig:wavelength-dependent-bias}
\end{figure}

\begin{figure}[h!tb]
    \centering
    \includegraphics[width=\textwidth]{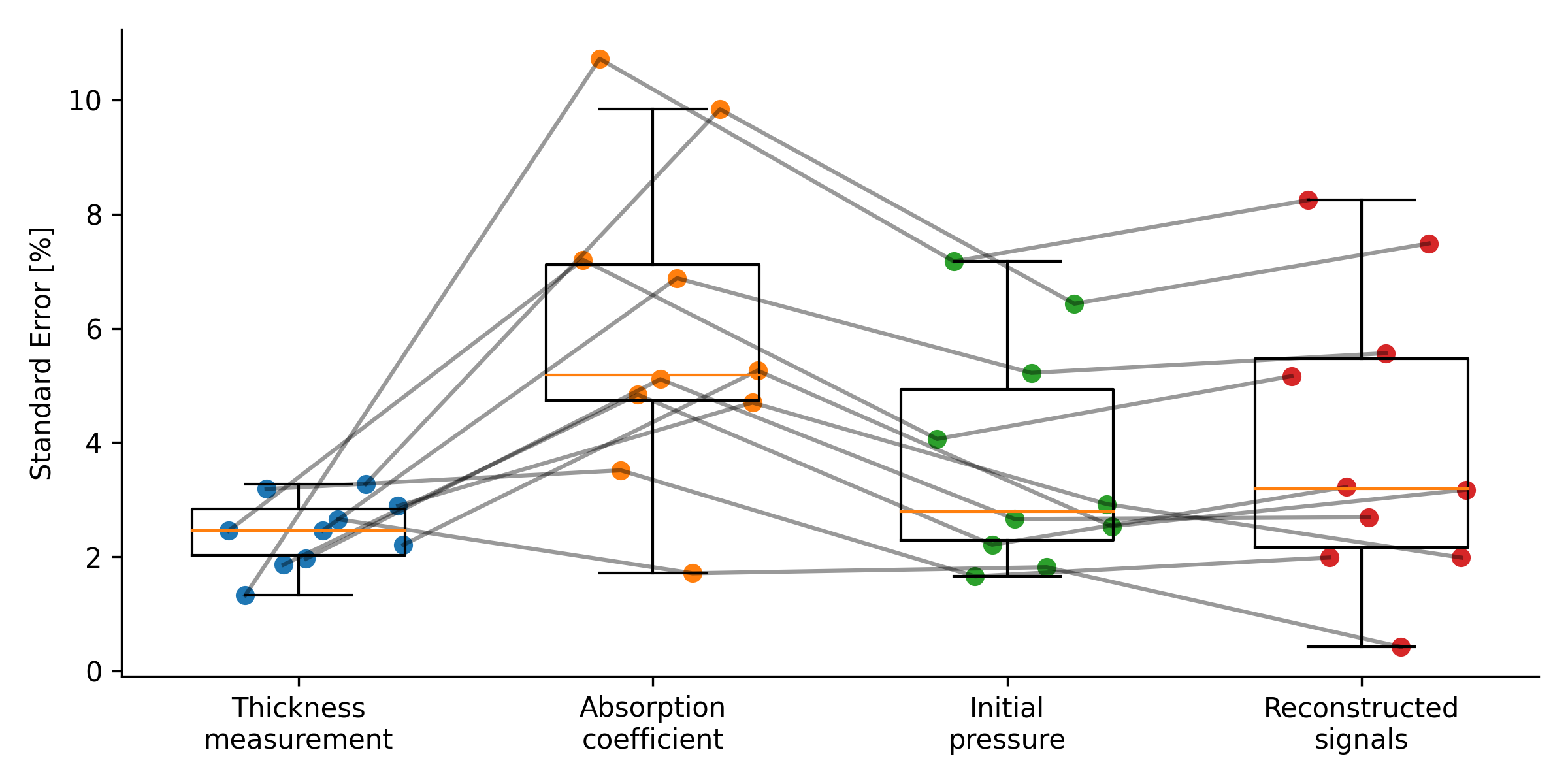}
    \caption{\textbf{The standard error propagating through various stages of the simulation process.} It should be noted that the standard errors for the initial pressure and reconstructions are estimates based on low-resolution simulations with the pipeline described in the paper.}
    \label{fig:error_propagation_from_thickness_measurements}
\end{figure}

\clearpage

\section*{Detailed Table of all phantom properties}

\begin{longtable}[h!tb]{cccc}
\caption{Tabulated overview of the characteristic optical properties of the tissue-mimicking phantoms at the reference wavelength of 800\,nm. Shown are the median results and the standard deviations for the optical absorption coefficient $\mu_a$ and the reduced optical scattering coefficient $\mu_s'$ of the 10 double-integrating sphere measurements made for each sample.}
\label{tab:my_label}\\
\textbf{Phantom Identifier} & \textbf{\quad\quad \quad Region\quad\quad\quad} & \quad\quad\textbf{$\boldsymbol{\mu_a}$ [cm$^{-1}$]}\quad\quad & \quad\quad\textbf{$\boldsymbol{\mu_s'}$ [cm$^{-1}$]}\quad\quad\\
\hline
\\
\textbf{P.5.32} & Background & 0.35 $\pm$ 0.013 & 8.86 $\pm$ 0.12 \\
 & Inclusion 1 & 3.36 $\pm$ 0.048 & 8.86 $\pm$ 0.16 \\
\textbf{P.5.31} & Background & 0.35 $\pm$ 0.013 & 8.86 $\pm$ 0.12 \\
 & Inclusion 1 & 3.36 $\pm$ 0.048 & 8.86 $\pm$ 0.16 \\
\textbf{P.5.29} & Background & 0.16 $\pm$ 0.012 & 7.53 $\pm$ 0.07 \\
 & Inclusion 1 & 1.47 $\pm$ 0.042 & 9.70 $\pm$ 0.27 \\
\textbf{P.5.27} & Background & 0.10 $\pm$ 0.014 & 12.27 $\pm$ 0.33 \\
 & Inclusion 1 & 1.47 $\pm$ 0.042 & 9.70 $\pm$ 0.27 \\
\textbf{P.5.25} & Background & 0.09 $\pm$ 0.016 & 8.64 $\pm$ 0.18 \\
 & Inclusion 1 & 1.47 $\pm$ 0.042 & 9.70 $\pm$ 0.27 \\
\textbf{P.5.10.2} & Background & 0.48 $\pm$ 0.021 & 7.27 $\pm$ 0.15 \\
 & Inclusion 1 & 1.11 $\pm$ 0.052 & 9.70 $\pm$ 0.32 \\
 & Inclusion 2 & 3.17 $\pm$ 0.158 & 5.34 $\pm$ 0.18 \\
\textbf{P.5.10.3} & Background & 0.48 $\pm$ 0.021 & 7.27 $\pm$ 0.15 \\
 & Inclusion 1 & 2.14 $\pm$ 0.037 & 10.03 $\pm$ 0.16 \\
\textbf{P.5.8.2} & Background & 0.39 $\pm$ 0.015 & 11.70 $\pm$ 0.25 \\
 & Inclusion 1 & 3.17 $\pm$ 0.158 & 5.34 $\pm$ 0.18 \\
 & Inclusion 2 & 2.14 $\pm$ 0.037 & 10.03 $\pm$ 0.16 \\
\textbf{P.5.8.3} & Background & 0.39 $\pm$ 0.015 & 11.70 $\pm$ 0.25 \\
 & Inclusion 1 & 1.11 $\pm$ 0.052 & 9.70 $\pm$ 0.32 \\
\textbf{P.5.6.2} & Background & 0.27 $\pm$ 0.042 & 7.38 $\pm$ 0.32 \\
 & Inclusion 1 & 1.11 $\pm$ 0.052 & 9.70 $\pm$ 0.32 \\
 & Inclusion 2 & 3.17 $\pm$ 0.158 & 5.34 $\pm$ 0.18 \\
\textbf{P.5.6.3} & Background & 0.27 $\pm$ 0.042 & 7.38 $\pm$ 0.32 \\
 & Inclusion 1 & 3.17 $\pm$ 0.158 & 5.34 $\pm$ 0.18 \\
\textbf{P.5.4.2} & Background & 0.09 $\pm$ 0.008 & 11.35 $\pm$ 0.29 \\
 & Inclusion 1 & 2.14 $\pm$ 0.037 & 10.03 $\pm$ 0.16 \\
 & Inclusion 2 & 1.11 $\pm$ 0.052 & 9.70 $\pm$ 0.32 \\
\textbf{P.5.4.3} & Background & 0.09 $\pm$ 0.008 & 11.35 $\pm$ 0.29 \\
 & Inclusion 1 & 3.17 $\pm$ 0.158 & 5.34 $\pm$ 0.18 \\
\textbf{P.5.2.2} & Background & 0.07 $\pm$ 0.012 & 7.63 $\pm$ 0.15 \\
 & Inclusion 1 & 3.17 $\pm$ 0.158 & 5.34 $\pm$ 0.18 \\
\textbf{P.5.2.3} & Background & 0.07 $\pm$ 0.012 & 7.63 $\pm$ 0.15 \\
 & Inclusion 1 & 1.11 $\pm$ 0.052 & 9.70 $\pm$ 0.32 \\
\textbf{P.5.24} & Background & 0.26 $\pm$ 0.016 & 6.39 $\pm$ 0.12 \\
 & Inclusion 1 & 3.17 $\pm$ 0.158 & 5.34 $\pm$ 0.18 \\
\textbf{P.5.23} & Background & 0.26 $\pm$ 0.016 & 6.39 $\pm$ 0.12 \\
 & Inclusion 1 & 2.50 $\pm$ 0.054 & 8.49 $\pm$ 0.12 \\
\textbf{P.5.22} & Background & 0.14 $\pm$ 0.018 & 13.70 $\pm$ 0.34 \\
 & Inclusion 1 & 2.14 $\pm$ 0.037 & 10.03 $\pm$ 0.16 \\
\textbf{P.5.21} & Background & 0.14 $\pm$ 0.018 & 13.70 $\pm$ 0.34 \\
 & Inclusion 1 & 2.50 $\pm$ 0.054 & 8.49 $\pm$ 0.12 \\
\textbf{P.5.20} & Background & 0.07 $\pm$ 0.007 & 8.66 $\pm$ 0.07 \\
 & Inclusion 1 & 3.17 $\pm$ 0.158 & 5.34 $\pm$ 0.18 \\
\textbf{P.5.19} & Background & 0.07 $\pm$ 0.007 & 8.66 $\pm$ 0.07 \\
 & Inclusion 1 & 2.50 $\pm$ 0.054 & 8.49 $\pm$ 0.12 \\
\textbf{P.5.16} & Background & 0.09 $\pm$ 0.009 & 11.68 $\pm$ 0.20 \\
 & Inclusion 1 & 1.28 $\pm$ 0.021 & 9.47 $\pm$ 0.12 \\
\textbf{P.5.15} & Background & 0.18 $\pm$ 0.013 & 8.06 $\pm$ 0.15 \\
 & Inclusion 1 & 1.28 $\pm$ 0.021 & 9.47 $\pm$ 0.12 \\
\textbf{P.5.14} & Background & 0.18 $\pm$ 0.013 & 8.06 $\pm$ 0.15 \\
 & Inclusion 1 & 0.85 $\pm$ 0.036 & 9.62 $\pm$ 0.26 \\
\textbf{P.5.13} & Background & 0.09 $\pm$ 0.009 & 11.68 $\pm$ 0.20 \\
 & Inclusion 1 & 0.85 $\pm$ 0.036 & 9.62 $\pm$ 0.26 \\
\textbf{P.5.12} & Background & 0.54 $\pm$ 0.028 & 11.35 $\pm$ 0.42 \\
 & Inclusion 1 & 2.14 $\pm$ 0.037 & 10.03 $\pm$ 0.16 \\
 & Inclusion 2 & 2.14 $\pm$ 0.037 & 10.03 $\pm$ 0.16 \\
\textbf{P.5.11} & Background & 0.54 $\pm$ 0.028 & 11.35 $\pm$ 0.42 \\
\textbf{P.5.9} & Background & 0.51 $\pm$ 0.018 & 7.30 $\pm$ 0.14 \\
\textbf{P.5.7} & Background & 0.34 $\pm$ 0.018 & 11.21 $\pm$ 0.20 \\
\textbf{P.5.5} & Background & 0.27 $\pm$ 0.013 & 7.64 $\pm$ 0.10 \\
\textbf{P.5.3} & Background & 0.09 $\pm$ 0.010 & 11.13 $\pm$ 0.19 \\
\textbf{P.5.1} & Background & 0.08 $\pm$ 0.006 & 7.83 $\pm$ 0.08 \\
\textbf{P.3.78} & Background & 0.10 $\pm$ 0.013 & 12.43 $\pm$ 0.31 \\
 & Inclusion 1 & 3.09 $\pm$ 0.066 & 10.95 $\pm$ 0.28 \\
\textbf{P.3.76} & Background & 0.10 $\pm$ 0.013 & 12.43 $\pm$ 0.31 \\
 & Inclusion 1 & 0.58 $\pm$ 0.021 & 10.09 $\pm$ 0.35 \\
\textbf{P.3.74} & Background & 0.21 $\pm$ 0.005 & 13.33 $\pm$ 0.19 \\
 & Inclusion 1 & 2.04 $\pm$ 0.119 & 10.95 $\pm$ 0.51 \\
\textbf{P.3.72} & Background & 0.15 $\pm$ 0.009 & 11.83 $\pm$ 0.27 \\
 & Inclusion 1 & 2.04 $\pm$ 0.119 & 10.95 $\pm$ 0.51 \\
 & Inclusion 2 & 3.09 $\pm$ 0.066 & 10.95 $\pm$ 0.28 \\
\textbf{P.3.71} & Background & 0.15 $\pm$ 0.009 & 11.83 $\pm$ 0.27 \\
 & Inclusion 1 & 1.04 $\pm$ 0.099 & 10.45 $\pm$ 0.61 \\
 & Inclusion 2 & 1.04 $\pm$ 0.099 & 10.45 $\pm$ 0.61 \\
\textbf{P.3.69} & Background & 0.27 $\pm$ 0.007 & 13.09 $\pm$ 0.16 \\
 & Inclusion 1 & 3.09 $\pm$ 0.066 & 10.95 $\pm$ 0.28 \\
 & Inclusion 2 & 0.53 $\pm$ 0.016 & 9.49 $\pm$ 0.23 \\
\textbf{P.3.68} & Background & 0.28 $\pm$ 0.008 & 10.28 $\pm$ 0.16 \\
 & Inclusion 1 & 2.04 $\pm$ 0.119 & 10.95 $\pm$ 0.51 \\
 & Inclusion 2 & 2.04 $\pm$ 0.119 & 10.95 $\pm$ 0.51 \\
\textbf{P.3.67} & Background & 0.28 $\pm$ 0.008 & 10.28 $\pm$ 0.16 \\
 & Inclusion 1 & 3.09 $\pm$ 0.066 & 10.95 $\pm$ 0.28 \\
 & Inclusion 2 & 2.04 $\pm$ 0.119 & 10.95 $\pm$ 0.51 \\
\textbf{P.3.65} & Background & 0.19 $\pm$ 0.008 & 14.90 $\pm$ 0.35 \\
 & Inclusion 1 & 0.53 $\pm$ 0.016 & 9.49 $\pm$ 0.23 \\
 & Inclusion 2 & 1.04 $\pm$ 0.099 & 10.45 $\pm$ 0.61 \\
\textbf{P.3.64} & Background & 0.20 $\pm$ 0.007 & 11.62 $\pm$ 0.27 \\
 & Inclusion 1 & 3.09 $\pm$ 0.066 & 10.95 $\pm$ 0.28 \\
 & Inclusion 2 & 3.09 $\pm$ 0.066 & 10.95 $\pm$ 0.28 \\
\textbf{P.3.62} & Background & 0.18 $\pm$ 0.011 & 10.24 $\pm$ 0.25 \\
 & Inclusion 1 & 2.04 $\pm$ 0.119 & 10.95 $\pm$ 0.51 \\
 & Inclusion 2 & 1.04 $\pm$ 0.099 & 10.45 $\pm$ 0.61 \\
\textbf{P.3.61} & Background & 0.18 $\pm$ 0.011 & 10.24 $\pm$ 0.25 \\
 & Inclusion 1 & 0.53 $\pm$ 0.016 & 9.49 $\pm$ 0.23 \\
 & Inclusion 2 & 1.04 $\pm$ 0.099 & 10.45 $\pm$ 0.61 \\
\textbf{P.3.60} & Background & 0.13 $\pm$ 0.014 & 12.95 $\pm$ 0.33 \\
 & Inclusion 1 & 3.09 $\pm$ 0.066 & 10.95 $\pm$ 0.28 \\
 & Inclusion 2 & 0.53 $\pm$ 0.016 & 9.49 $\pm$ 0.23 \\
\textbf{P.3.59} & Background & 0.13 $\pm$ 0.014 & 12.95 $\pm$ 0.33 \\
 & Inclusion 1 & 0.53 $\pm$ 0.016 & 9.49 $\pm$ 0.23 \\
 & Inclusion 2 & 3.09 $\pm$ 0.066 & 10.95 $\pm$ 0.28 \\
\textbf{P.3.57} & Background & 0.12 $\pm$ 0.012 & 11.22 $\pm$ 0.25 \\
 & Inclusion 1 & 2.04 $\pm$ 0.119 & 10.95 $\pm$ 0.51 \\
 & Inclusion 2 & 1.04 $\pm$ 0.099 & 10.45 $\pm$ 0.61 \\
\textbf{P.3.55} & Background & 0.08 $\pm$ 0.005 & 16.24 $\pm$ 0.43 \\
 & Inclusion 1 & 0.53 $\pm$ 0.016 & 9.49 $\pm$ 0.23 \\
 & Inclusion 2 & 0.53 $\pm$ 0.016 & 9.49 $\pm$ 0.23 \\
\textbf{P.3.53} & Background & 0.09 $\pm$ 0.008 & 12.52 $\pm$ 0.18 \\
 & Inclusion 1 & 3.09 $\pm$ 0.066 & 10.95 $\pm$ 0.28 \\
 & Inclusion 2 & 2.04 $\pm$ 0.119 & 10.95 $\pm$ 0.51 \\
\textbf{P.3.52} & Background & 0.10 $\pm$ 0.012 & 9.41 $\pm$ 0.17 \\
 & Inclusion 1 & 3.09 $\pm$ 0.066 & 10.95 $\pm$ 0.28 \\
 & Inclusion 2 & 1.04 $\pm$ 0.099 & 10.45 $\pm$ 0.61 \\
\textbf{P.3.50} & Background & 0.44 $\pm$ 0.019 & 15.74 $\pm$ 0.47 \\
 & Inclusion 1 & 2.04 $\pm$ 0.119 & 10.95 $\pm$ 0.51 \\
\textbf{P.3.48} & Background & 0.44 $\pm$ 0.015 & 10.39 $\pm$ 0.20 \\
 & Inclusion 1 & 0.53 $\pm$ 0.016 & 9.49 $\pm$ 0.23 \\
\textbf{P.3.47} & Background & 0.41 $\pm$ 0.015 & 7.29 $\pm$ 0.19 \\
 & Inclusion 1 & 0.08 $\pm$ 0.017 & 9.78 $\pm$ 0.19 \\
\textbf{P.3.46} & Background & 0.34 $\pm$ 0.016 & 4.91 $\pm$ 0.11 \\
 & Inclusion 1 & 3.09 $\pm$ 0.066 & 10.95 $\pm$ 0.28 \\
\textbf{P.3.44} & Background & 0.20 $\pm$ 0.022 & 12.68 $\pm$ 0.29 \\
 & Inclusion 1 & 0.53 $\pm$ 0.016 & 9.49 $\pm$ 0.23 \\
\textbf{P.3.41} & Background & 0.16 $\pm$ 0.022 & 5.78 $\pm$ 0.14 \\
 & Inclusion 1 & 2.04 $\pm$ 0.119 & 10.95 $\pm$ 0.51 \\
\textbf{P.3.40} & Background & 0.09 $\pm$ 0.008 & 14.86 $\pm$ 0.23 \\
 & Inclusion 1 & 0.53 $\pm$ 0.016 & 9.49 $\pm$ 0.23 \\
\textbf{P.3.38} & Background & 0.12 $\pm$ 0.019 & 10.23 $\pm$ 0.25 \\
 & Inclusion 1 & 3.09 $\pm$ 0.066 & 10.95 $\pm$ 0.28 \\
\textbf{P.3.37} & Background & 0.05 $\pm$ 0.009 & 8.05 $\pm$ 0.12 \\
 & Inclusion 1 & 2.04 $\pm$ 0.119 & 10.95 $\pm$ 0.51 \\
\textbf{P.3.36} & Background & 0.05 $\pm$ 0.013 & 5.35 $\pm$ 0.09 \\
 & Inclusion 1 & 1.04 $\pm$ 0.099 & 10.45 $\pm$ 0.61 \\
\textbf{P.3.35} & Background & 0.07 $\pm$ 0.008 & 12.91 $\pm$ 0.25 \\
 & Inclusion 1 & 1.04 $\pm$ 0.099 & 10.45 $\pm$ 0.61 \\
\textbf{P.3.34} & Background & 0.05 $\pm$ 0.013 & 10.24 $\pm$ 0.26 \\
 & Inclusion 1 & 3.09 $\pm$ 0.066 & 10.95 $\pm$ 0.28 \\
\textbf{P.3.32} & Background & 0.01 $\pm$ 0.008 & 7.34 $\pm$ 0.08 \\
 & Inclusion 1 & 1.04 $\pm$ 0.099 & 10.45 $\pm$ 0.61 \\
\textbf{P.3.31} & Background & 0.02 $\pm$ 0.003 & 5.89 $\pm$ 0.03 \\
 & Inclusion 1 & 0.53 $\pm$ 0.016 & 9.49 $\pm$ 0.23 \\
\textbf{P.3.30} & Background & 0.01 $\pm$ 0.002 & 15.35 $\pm$ 0.32 \\
 & Inclusion 1 & 3.09 $\pm$ 0.066 & 10.95 $\pm$ 0.28 \\
\textbf{P.3.28} & Background & 0.01 $\pm$ 0.000 & 9.40 $\pm$ 0.00 \\
 & Inclusion 1 & 1.04 $\pm$ 0.099 & 10.45 $\pm$ 0.61 \\
\textbf{P.3.27} & Background & 0.00 $\pm$ 0.000 & 7.59 $\pm$ 0.00 \\
 & Inclusion 1 & 0.53 $\pm$ 0.016 & 9.49 $\pm$ 0.23 \\
\textbf{P.3.26} & Background & 0.00 $\pm$ 0.000 & 5.25 $\pm$ 0.00 \\
 & Inclusion 1 & 0.08 $\pm$ 0.017 & 9.78 $\pm$ 0.19 \\
\textbf{P.3.25} & Background & 0.44 $\pm$ 0.019 & 15.74 $\pm$ 0.47 \\
\textbf{P.3.24} & Background & 0.42 $\pm$ 0.009 & 12.66 $\pm$ 0.22 \\
\textbf{P.3.23} & Background & 0.44 $\pm$ 0.015 & 10.39 $\pm$ 0.20 \\
\textbf{P.3.22} & Background & 0.41 $\pm$ 0.015 & 7.29 $\pm$ 0.19 \\
\textbf{P.3.21} & Background & 0.34 $\pm$ 0.016 & 4.91 $\pm$ 0.11 \\
\textbf{P.3.20} & Background & 0.20 $\pm$ 0.008 & 15.30 $\pm$ 0.53 \\
\textbf{P.3.19} & Background & 0.20 $\pm$ 0.022 & 12.68 $\pm$ 0.29 \\
\textbf{P.3.18} & Background & 0.18 $\pm$ 0.016 & 10.21 $\pm$ 0.32 \\
\textbf{P.3.17} & Background & 0.15 $\pm$ 0.008 & 8.49 $\pm$ 0.17 \\
\textbf{P.3.16} & Background & 0.16 $\pm$ 0.022 & 5.78 $\pm$ 0.14 \\
\textbf{P.3.15} & Background & 0.09 $\pm$ 0.008 & 14.86 $\pm$ 0.23 \\
\textbf{P.3.14} & Background & 0.10 $\pm$ 0.009 & 11.61 $\pm$ 0.31 \\
\textbf{P.3.13} & Background & 0.12 $\pm$ 0.019 & 10.23 $\pm$ 0.25 \\
\textbf{P.3.12} & Background & 0.05 $\pm$ 0.009 & 8.05 $\pm$ 0.12 \\
\textbf{P.3.11} & Background & 0.05 $\pm$ 0.013 & 5.35 $\pm$ 0.09 \\
\textbf{P.3.10} & Background & 0.07 $\pm$ 0.008 & 12.91 $\pm$ 0.25 \\
\textbf{P.3.9} & Background & 0.05 $\pm$ 0.013 & 10.24 $\pm$ 0.26 \\
\textbf{P.3.8} & Background & 0.03 $\pm$ 0.006 & 9.54 $\pm$ 0.20 \\
\textbf{P.3.7} & Background & 0.01 $\pm$ 0.008 & 7.34 $\pm$ 0.08 \\
\textbf{P.3.6} & Background & 0.02 $\pm$ 0.003 & 5.89 $\pm$ 0.03 \\
\textbf{P.3.5} & Background & 0.01 $\pm$ 0.002 & 15.35 $\pm$ 0.32 \\
\textbf{P.3.4} & Background & 0.00 $\pm$ 0.002 & 12.76 $\pm$ 0.12 \\
\textbf{P.3.3} & Background & 0.01 $\pm$ 0.000 & 9.40 $\pm$ 0.00 \\
\textbf{P.3.2} & Background & 0.00 $\pm$ 0.000 & 7.59 $\pm$ 0.00 \\
\textbf{P.3.1} & Background & 0.00 $\pm$ 0.000 & 5.25 $\pm$ 0.00 \\
\textbf{P.2.4} & Background & 0.08 $\pm$ 0.013 & 17.64 $\pm$ 0.73 \\
 & Inclusion 1 & 1.41 $\pm$ 0.090 & 10.66 $\pm$ 0.48 \\
\textbf{P.2.3} & Background & 0.15 $\pm$ 0.030 & 10.06 $\pm$ 0.25 \\
 & Inclusion 1 & 0.85 $\pm$ 0.021 & 10.63 $\pm$ 0.21 \\
\textbf{P.2.2} & Background & 0.08 $\pm$ 0.013 & 17.64 $\pm$ 0.73 \\
 & Inclusion 1 & 0.85 $\pm$ 0.021 & 10.63 $\pm$ 0.21 \\
\textbf{P.2.1} & Background & 0.15 $\pm$ 0.030 & 10.06 $\pm$ 0.25 \\
 & Inclusion 1 & 1.41 $\pm$ 0.090 & 10.66 $\pm$ 0.48 \\
\textbf{P.1.11} & Background & 0.18 $\pm$ 0.063 & 15.25 $\pm$ 1.45 \\
 & Inclusion 1 & 0.85 $\pm$ 0.021 & 10.63 $\pm$ 0.21 \\
 & Inclusion 2 & 0.85 $\pm$ 0.021 & 10.63 $\pm$ 0.21 \\
 & Inclusion 3 & 1.41 $\pm$ 0.090 & 10.66 $\pm$ 0.48 \\
 & Inclusion 4 & 1.41 $\pm$ 0.090 & 10.66 $\pm$ 0.48 \\
\textbf{P.1.6} & Background & 0.12 $\pm$ 0.009 & 10.54 $\pm$ 0.28 \\
 & Inclusion 1 & 1.14 $\pm$ 0.023 & 8.06 $\pm$ 0.14 \\
 & Inclusion 2 & 1.14 $\pm$ 0.023 & 8.06 $\pm$ 0.14 \\
\textbf{P.1.3.a} & Background & 0.24 $\pm$ 0.023 & 5.36 $\pm$ 0.14 \\
\textbf{P.1.2.a} & Background & 0.13 $\pm$ 0.026 & 3.81 $\pm$ 0.17 \\
\textbf{P.1.1.a} & Background & 0.03 $\pm$ 0.021 & 1.49 $\pm$ 0.04 \\
\end{longtable}

\end{document}